\documentclass[superscriptaddress,amsmath,amssymb, aps, pra]{revtex4-2}

\usepackage{graphicx}
\usepackage{dcolumn}
\usepackage{bm}
\usepackage[colorlinks]{hyperref}
\usepackage[T1]{fontenc}
\usepackage[dvipsnames]{xcolor}
\usepackage{comment}

\newcommand{\bigzero}{\mbox{\normalfont\Large\bfseries $\mathbb{I}$}}
\newcommand{\FirstLayer}{\mbox{\normalfont\Large\bfseries $A^{(1)}$}}
\newcommand{\SecondLayer}{\mbox{\normalfont\Large\bfseries $A^{(2)}$}}
\newcommand{\rvline}{\hspace*{-\arraycolsep}\vline\hspace*{-\arraycolsep}}

\begin{document}

\title{Discrete-time Quantum Walk on Multilayer Networks}

\author{M. N. Jayakody}
\affiliation{Faculty of Engineering and the Institute of Nanotechnology and Advanced Materials, Bar-Ilan University, Ramat Gan 5290002, Israel}

\author{Priodyuti Pradhan}
\affiliation{Department of Computer Science and Engineering, Indian Institute of Information Technology Raichur, Karnataka - 584135, India}

\author{Dana Ben Porath}
\affiliation{Faculty of Engineering and the Institute of Nanotechnology and Advanced Materials, Bar-Ilan University, Ramat Gan 5290002, Israel}

\author{E. Cohen}
\affiliation{Faculty of Engineering and the Institute of Nanotechnology and Advanced Materials, Bar-Ilan University, Ramat Gan 5290002, Israel}

\begin{abstract}
Multilayer network is a potent platform which paves a way to study the interactions among entities in various networks with multiple types of relationships. In this study, the dynamics of discrete-time quantum walk on a multilayer network are explored in detail. We derive recurrence formulae for the coefficients of the wave function of a quantum walker on an undirected graph with finite number of nodes. By extending these formulae to include extra layers, we develop a simulation model to describe the time-evolution of the quantum walker on a multilayer network. The time-averaged probability and the return probability of the quantum walker are studied in relation to Fourier and Grover walks on multilayer networks. Furthermore, we analyze the impact of decoherence on the quantum transport, shedding light on how environmental interactions may impact the behavior of quantum walkers on multilayer network structures.
\end{abstract}

\maketitle

\section{Introduction}

Quantum walks (QWs) are the quantum analogues of classical random walks (CRWs). Importantly, QWs contribute to theoretical and applied studies of quantum computing \cite{Childs_2013} in general and quantum algorithms \cite{Kempe} in particular. There are two broad classes of QWs known as discrete-time QWs (DTQW) and continuous-time QWs (CTQW), each of which has significant distinctions in their mathematical formalism \cite{Wang}. Considerable body of work can be found in literature which explore the dynamics of linear and cyclic QWs in two and higher dimensional spaces, as well as on specific graphs \cite{venegas2012quantum,Wang}. 

In the study of complex systems, multilayer networks play a crucial  role as a modeling tool \cite{aleta2019multilayer,jalan2018localization}. A multilayer network consists of nodes and edges, yet the nodes exist in separate layers representing different forms of interactions. Multilayer networks are used to describe the behavior patterns and evolution of ecological systems \cite{pilosof2017multilayer}, complex interactions across multiple layers of biological systems \cite{gosak2018network}, public transportation systems \cite{gallotti2015multilayer,gallotti2014anatomy} and the structure of financial
markets \cite{musmeci2017multiplex}. In literature, one can find several studies which utilize the framework of CTQWs to study the transport properties of multilayer dendrimer networks \cite{galiceanu2016continuous}, honeycomb networks \cite{maquine2022quantum} and scale-free networks \cite{maciel2020quantum} in detail. Nonetheless, it is clear that there has been insufficient exploration of the dynamics of DTQWs on multilayer networks. Therefore, to address this gap, we present a comprehensive study of DTQWs on a multilayer network. The most general form of networks and multilayer networks are mathematically represented using the notion of a graph \cite{aleta2019multilayer}. Hence, its worth exploring the research work on DTQWs that incorporate graphs. Several formulations of the DTQWs on specific graph structures can be found in literature. However, defining a DTQWs on an arbitrary graph is more difficult than that of a CTQWs \cite{ambainis2003quantum}. Ref. \cite{aharonov2001quantum} has proposed a framework for defining QWs on regular graphs. Sometimes this framework is termed as Shunt-Decomposition model \cite{godsil2019discrete}. In \cite{Kempe}, it is suggested to add one or more self-loops to each vertex with the purpose of obtaining a regular graphs when modeling the QWs on irregular graphs. Extending the idea given in \cite{watrous2001quantum}, Kendon \cite{kendon2006quantum} has presented a method to simulate DTQWs on general undirected graphs. Sometimes this method is identified as Arc-Reversal model \cite{godsil2019discrete}. Ref. \cite{feldman2004scattering} has proposed another framework for QWs on a graph in which the motion of the quantum walker takes place on the edges of the graph rather than the vertices. 

In our work, we give a block matrix representation for the state of the quantum walker instead of the usual column matrix representation. Then, we derive recurrence formulae that imitate the coin toss and the shifting operation of the QW. We adopt the shift operation given in Arc-Reversal model to develop our framework. Using the recurrence formulae and the block matrix representation, we develop a simulation to mimic the progression of QWs on an undirected graph. Later we extend our framework to mimic QWs on multilayer networks. The paper is organized in the following way. In section \ref{QWs on a graph}, we present our mathematical model for QWs on a graph. Section \ref{QW_on_multilayer} is dedicated to extend our mathematical model to include QWs on multilayer networks. Moreover, for the sake of comparison purposes, we model a CRW on a multilayer network as well in section \ref{CRW_on_MLN}. Numerical implementation of QWs on a toy model and some synthetic multilayer networks are given in section \ref{Analysis 1} and \ref{Analysis 2}, respectively, along with a detailed analysis on time-average probability, return probability and decoherence.

\section{QWs on a Graph}\label{QWs on a graph}
Consider a finite undirected graph $\mathcal{G} =\{V,E\}$ where $V=\{v_1,\ldots,v_n\}$ is the set of vertices (nodes) and $E=\{(v_i, v_j) \ | \ v_i,v_j \in V \}$ is the set of edges (connections). Note that, the graphs which are studied here are finite as opposed to the unrestricted line or the integer line we used to define the QWs on a line. In our study, we consider simple graphs (i.e. graphs without self-loops or parallel edges). We denote the adjacency matrix corresponding to $\mathcal{G}$ as ${\bf A} \in \mathbb{R}^{n \times n}$ which is defined in the following way 
\begin{equation}\label{adjacency_matrix}
a_{ij}=
   \begin{cases}
     1       & \quad \text{If} \ (v_i, v_j)\in E \\
     0  & \quad \text{Otherwise}\\
   \end{cases}
\end{equation}
The positive integers $n=|V|$ and $m=|E|$ represent the number of nodes and number of edges in $\mathcal{G}$, respectively. The number of edges linked to a particular node $v_i$ is referred to as its degree and denoted by $d_i = \sum_{j=1}^{n} a_{ij}$. Let us now model the propagation of a QW on the graph $\mathcal{G}$. For the sake of convenience, let us relabel the vertices of the graph or in QWs' terminology, the position states of the QW as $|x\rangle_p$ where $x=1,\ldots,n$. Then, the set of vertices $V=\{|1\rangle_p,|2\rangle_p,\ldots,|n\rangle_p\}$ becomes the position basis set of the QW which spans the position Hilbert space $H_{p}$. Let us denote the subspace spanned by the basis element $|x\rangle_p$ in $V$ as $H_{p}^{(x)}$. That is, $H_{p}^{(x)}=\text{span}\{|x\rangle_p\}$ where $|x\rangle_p \in V$. Now, for each vertex $|x\rangle_p$, let us assign a coin Hilbert space $H_{c}^{(x)}$ spanned by the coin basis $\{|r\rangle_c \ | \ r=1, \hdots, d_x\}$ where $d_{x}$ is the degree of the vertex $|x\rangle_p$. Note that, the dimension of the coin Hilbert space $H_{c}^{(x)}$ is $d_{x}$. For the purpose of gaining an advantage in simulation, we adopt the following strategy to modify the labeling of the coin states. We define a set $\mathcal{B}_{x}$ that comprises the labels of the vertices adjacent to the vertex $|x\rangle_p$ as $\mathcal{B}_{x}=\{y \  :\  \text{vertex} \ |x\rangle_p \ \text{and} \ |y\rangle_p  \ \text{have a connection}  \}$. Note that, $|\mathcal{B}_{x}|=d_{x}$. Now, define a function as, $f_{x}(r)= r^{th} \text{ element of} \ \mathcal{B}_{x}$. Since parallel edges are excluded in our study, the function $f_{x}(r)$ is a bijection. Without the loss of generality, we always arrange the elements of $\mathcal{B}_{x}$ in the ascending order. Now, we can denote the basis of the coin Hilbert space $H_c^{(x)}$ as $\{|f_{x}(r)\rangle_c \ | \ r=1, \hdots, d_{x}\}$. Note that, by using the function $f_{x}(r)$, we have labeled the coin states of each vertex using the edges connected to it. Such an approach can be found in the study \cite{watrous2001quantum}. Now, the state vector of the quantum walker at position $|x\rangle_p$ at time $t$ can be written as
\begin{equation}\label{State_vector_at_x_at_t}
    |\psi(x,t)\rangle=\sum_{r=1}^{d_{x}} \alpha_{x,f_x(r)}(t)|x\rangle_p| f_x(r) \rangle_c
\end{equation} 
where $\alpha_{x,f_x(r)}(t) \in \mathbb{C}$ are called the probability amplitudes, $|x\rangle_p \in H_p^{(x)}$, $| f_x(r) \rangle_c \in H_c^{(x)}$ and $|\psi(x,t)\rangle \in H_p^{(x)} \otimes H_c^{(x)} $. Note that, $dim(H_p^{(x)})=1$, $dim(H_c^{(x)})=d_x$ and $dim(H_p^{(x)} \otimes H_c^{(x)})=d_x$. For each position $x$, the coefficients of $\alpha_{x,f_x(r)}(t)$ are defined as follows
\begin{equation}\label{coeffienct_Eq}
\alpha_{x,r}(t)=
   \begin{cases}
     \alpha_{x,r}(t) & \quad \text{If} \ r \in \mathcal{B}_{x} \\
     0  & \quad \text{Otherwise}\\
   \end{cases}
\end{equation}
Our next task is to write an expression for the total wave function of the quantum walker on $\mathcal{G}$ at time $t$. For that, we need to sum the state vectors $|\psi(x,t)\rangle$ in \eqref{State_vector_at_x_at_t} over all the vertices. However, we are unable to perform such a summation because the state vectors $|\psi(x,t)\rangle$ corresponding to each vertex resides in different composite Hilbert spaces. Note that, the size of the coin Hilbert space $H_c^{(x)}$ changes with the degree of the node $x$. Therefore, to perform such a summation, one needs to combine the set of composite Hilbert spaces $\{H_p^{(x)} \otimes H_c^{(x)}\}_x$ in a reasonable manner, with the purpose of forming a bigger Hilbert space that includes all the state vectors. The operation of direct sum of vector spaces paves a way to combine the composite Hilbert spaces to cater our demand. Let us define $H=\bigoplus \limits_{x=1}^{n}\bigg(H_p^{(x)} \otimes H_c^{(x)}\bigg)$ where $\oplus$ denotes the external direct sum of Hilbert spaces and $\textit{dim}(H)=\sum_{x=1}^{n}d_x$. According to the definition of the global Hilbert space $H$, it is obvious that for each vertex $x$, the state vector $|\psi(x,t)\rangle \in H$. Hence, the total wave function of the quantum walker at time $t$ can be calculated by summing $|\psi(x,t)\rangle$ in \eqref{State_vector_at_x_at_t} over all the vertices. Then, the total wave function at time $t$ can be written as
\begin{equation}\label{State_vector_at_t}
|\psi_t\rangle=\sum_{x=1}^{n}\sum_{r=1}^{d_x} \alpha_{x,f_x(r)}(t)|x\rangle_p| f_x(r) \rangle_c
\end{equation} 
where $|\psi_t\rangle \in H$ and $\sum_{x=1}^{n}\sum_{r=1}^{d_x}|\alpha_{x,f_x(r)}(t)|^2=1$. Coin operator $C^{(x)}$ which acts on the coin states associated to the vertex $|x\rangle_p$ holds the transition probabilities from $|x\rangle_p$ to its neighbouring vertices. Hence, $C^{(x)}$ can be defined as
\begin{equation}\label{coin_opertor_at_x}
C^{(x)}=\sum_{i=1}^{d_x}\sum_{j=1}^{d_x}C^{(x)}_{ij}|f_x(i) \rangle \langle f_x(j)|
\end{equation} 
where $\{|f_x(r) \rangle_c\}_{r=1}^{d_x}$ are the basis elements of $H_c^{(x)}$. The coin coefficients $C^{(x)}_{ij} \in \mathbb{C}$ are chosen in such a way that the condition of unitarity of the QW is preserved, i.e. the total probability is unity at all time steps. Hence, $(C^{(x)})^{\dagger}C^{(x)}=C^{(x)}(C^{(x)})^{\dagger}=\mathbb{I}$. By combining each local coin operator $C^{(x)}$, one can write the global coin operator $C$ acting on $|\psi_t\rangle \in H$ as follows 
\begin{equation}\label{coin_opertor}
C=\sum_{x=1}^{n}\sum_{i=1}^{d_x}\sum_{j=1}^{d_x}C^{(x)}_{ij}|x \rangle \langle x| \otimes |f_x(i) \rangle \langle f_x(j)|
\end{equation} 
The shift operator of the QW on a graph is defined as follows
\begin{equation}\label{shift_opertor}
S|x \rangle_p |y \rangle_c = |y \rangle_p |x \rangle_c 
\end{equation}
Note that, both $C$ and $S$ are unitary operators associated to the Hilbert space $H$. Hence, a single step progression of the quantum walker on the graph is given by
\begin{equation}\label{single_step_progrsiion}
    |\psi_{t+1} \rangle=U|\psi_{t} \rangle
\end{equation}
where $U=SC$ is the evolution operator and $S$ and $C$ are the shift and coin operators respectively. 

\subsection{Matrix representation and simulation}
The local coin operator $C^{(x)}$, associated to the vertex $|x \rangle_p$ holds the transition probabilities from $|x \rangle_p$ to its neighbouring nodes. Suppose the vertex $|x \rangle_p$ is linked to $d_x$ number of vertices denoted by $|y_1\rangle_p, \hdots , |y_{d_x}\rangle_p$. Moreover, suppose that $y_1< \hdots < y_{d_x}$. Hence, for each $r \in \{1,\hdots , d_x\}$, we can write $y_r=f_x(r)$. Then, the block matrix of $C^{(x)}$ can be written as
\begin{equation}\label{Coin_opertor_at_x}
 C^{(x)}= \bordermatrix{ & \langle f_x(1)| & \hdots & \langle f_x(d_x)| \cr 
  |f_x(1)\rangle     & C^{(x)}_{11}  & \hdots & C^{(x)}_{1d_{x}} \cr 
  \vdots        & \vdots & \vdots & \vdots \cr
  |f_x(d_x)\rangle     & C^{(x)}_{d_{x}1}  & \hdots & C^{(x)}_{d_{x}d_{x}} \cr }
\end{equation}
As an example, the local coin operators associated to a four vertex graph is shown in Appendix \ref{App1}. In standard mathematical formalism of QWs \cite{jayakody2021one}, the state of the quantum walker at time $t$ is represented by a column vector. However, alternatively, one can give a convenient block matrix representation for the total wave function of the quantum walker on a graph given in \eqref{State_vector_at_t} as follows
\begin{equation}\label{block_matrix_A}
 |\psi_t\rangle={\bf N}_t= \bordermatrix{ & |1 \rangle_c   & \hdots & |n \rangle_c  \cr 
  |1\rangle_p   & \alpha_{1,1}(t)  & \hdots & \alpha_{1,n}(t) \cr 
  \vdots        & \vdots           & \vdots & \vdots \cr
  |n\rangle_p   & \alpha_{n,1}(t)  & \hdots & \alpha_{n,n}(t) \cr}
\end{equation}
In matrix  ${\bf N}_t$, the rows represent the position states and columns represent the coin states. Single row holds the coefficients corresponding to the coin states associated to a single position. According to the definition of the coefficients of $\alpha_{x,r}(t)$, given in \eqref{coeffienct_Eq}, some entries of the matrix ${\bf N}_t$ become zero (see Appendix \ref{App1}). Such a block matrix representation of the total wave function can be found in the study \cite{manouchehri2009quantum}. Note that, one can view the block matrix representation given in \eqref{block_matrix_A} as an adjacency matrix of a weighted graph having time dependent complex weights. By knowing all the coefficients of $\alpha_{x,r}(t)$, in other words, all the elements of ${\bf N}_t$, we can uniquely determine the total wave function of the quantum walker on the graph at time $t$. In addition, by updating the elements of ${\bf N}_t$ appropriately, one can determine the matrix ${\bf N}_{t+1}$ and hence the total wave function of the quantum walker at time $t+1$. The elements of ${\bf N}_t$ are updated in two processes. First, an intermediate matrix ${\bf \widetilde{N}}_t$ is generated using the following formula 
\begin{equation}\label{update_rule1}
\tilde{\alpha}_{x,f_x(i)}(t)=\sum_{j=1}^{d_x}{\alpha_{x,f_x(j)}(t)C^{(x)}_{ij}}
\end{equation}
Afterwards, the matrix ${\bf N}_{t+1}$ is determined by
taking the transpose of ${\bf \widetilde{N}}_t$. The update rule is given by
\begin{equation}\label{update_rule2}
\alpha_{x,f_x(i)}(t+1)=\tilde{\alpha}_{f_x(i),x}(t)
\end{equation}
Note that, the formulas given in \eqref{update_rule1} and \eqref{update_rule2} correspond to the coin flip and shift operation of the QW on the graph. Proof is given in the Appendix \ref{App2}. 
\subsection{Probability Calculation}
The probability $P_q(x,t)$ of finding the quantum walker at vertex $x$ at time $t$ can be calculated using \eqref{State_vector_at_x_at_t} as follows
\begin{equation}\label{prob_at_x}
        P_q(x,t)=\sum_{r=1}^{d_x}|\alpha_{x,f_x(r)}(t)|^2
\end{equation}
Note that, $P_q(x,t)$ can be determined by summing the elements in the $x^{th}$ row of the matrix ${\bf N}_t \odot {\bf N}_t^{*}$ where $\odot$ is the Hadamard product of matrices and ${\bf N}_t^{*}$ is the complex conjugate of the block matrix given in \eqref{block_matrix_A}.
\section{QWs on Multilayer networks} \label{QW_on_multilayer}
A multilayer network is a pair $\mathcal{M}=(\mathcal{G},\mathcal{C})$ where $\mathcal{G}=\{\mathcal{L}_{\alpha};\; \alpha \in \{1,2,\ldots, l\}\}$ is a family of undirected graphs of $\mathcal{L}_{\alpha} =\{V_{\alpha},E_{\alpha}\}$ (called layers of $\mathcal{M}$) with $V_{\alpha}=\{v_1^{\alpha}, v_2^{\alpha},\ldots,v_{n_{\alpha}}^{\alpha}\}$ is the set of vertices and $E_{\alpha}=\{e_1^{\alpha}, e_2^{\alpha},\ldots,e_{r_{\alpha}}^{\alpha}:e_{r_{\alpha}}=(v_i^{\alpha}, v_j^{\alpha})\} $ is the set of edges in the $\mathcal{L}_{\alpha}$ layer of the multilayer network \cite{jalan2018localization}. The positive integers $l$ , $n_{\alpha}$ and $r_{\alpha}$ are termed as the number of layers in $\mathcal{M}$, number of vertices and edges of the layer $\mathcal{L}_{\alpha}$ respectively. Moreover, $\mathcal{C}=\{E_{\alpha\beta}\subseteq V_{\alpha} \times V_{\beta}:\; \alpha, \beta \in \{1,2,\ldots, l \}, \alpha \neq \beta \}$ is the set of edges between $\mathcal{L}_\alpha$ and $\mathcal{L}_\beta$ layers. 
The elements of $\mathcal{C}$ are called crossed layers. Further, the elements of each $E_{\alpha}$ are called the set of intralayer edges and the elements of each $E_{\alpha\beta}$ ($\alpha \neq \beta$) are called the interlayer edges of $\mathcal{M}$ \cite{boccaletti2014structure}. Let us denote the adjacency matrices corresponding to each layer $\mathcal{L}_{\alpha}$ as $A^{(\alpha)} =(a^{\alpha}_{ij})\in \mathbb{R}^{n_{\alpha} \times n_{\alpha}}$ which is defined by 
\begin{equation}\label{adjacency matrices_MN}
a^{\alpha}_{ij}=
   \begin{cases}
     1       & \quad \text{If} \ (v_i^{\alpha}, v_j^{\alpha})\in E_{\alpha} \\
     0  & \quad \text{Otherwise}
   \end{cases} 
\end{equation}
for $1 \leq i,j \leq n_{\alpha}$ and $1 \leq \alpha \leq l$ where $n_{\alpha}$ is the number of nodes in layer $\mathcal{L}_\alpha$. The inter layer adjacency matrix corresponding to $E_{\alpha\beta}$ is the matrix $A^{[\alpha,\beta]} \in \mathbb{R}^{n_{\alpha} \times n_{\beta}}$ given by 
\begin{equation}\label{inter layer adjacency matrix}
a^{\alpha \beta}_{ij}=
   \begin{cases}
     1       & \quad \text{If} \ (v_i^{\alpha}, v_j^{\beta})\in E_{\alpha\beta} \\
     0  & \quad \text{Otherwise}
   \end{cases} 
\end{equation}
By combining the adjacency matrices corresponding to each layer $\mathcal{L}_{\alpha}$ and the inter-layer adjacency matrices appropriately, one can derive a supra-adjacency matrix \cite{boccaletti2014structure} which characterize the multilayer network $\mathcal{M}$. 

Now let us define a QW on the multilayer network structure. Recall that, in section \ref{QWs on a graph}, we developed a mathematical model to mimic the propagation of a QW on any given undirected graph. In our model, we acquire the information of the graphical structure by using the adjacency matrix of the graph. According to our model, when the adjacency matrix is given, we define the sets of $\{\mathcal{B}_{x}\}_{x=1}^{n}$ along with the set of functions $\{f_x(r)\}_{x=1}^{n}$ and then simulate the evolution of the QW on the graph by updating the elements of the block matrix in \eqref{block_matrix_A}. Likewise, one can use the same mathematical model to mimic the propagation of a quantum walker on a multilayer network just by following the same procedure given in section \ref{QWs on a graph} with the supra-adjacency matrix of the multilayer network. In a QW, the transition probabilities from vertex $x$ to its neighbouring vertices are given by the coin operator $C^{(x)}$ attached to the vertex $x$. One can choose suitable coin operators according to the simulation to control the probability flow from one vertex to the neighbouring vertices in the multilayer network. When the transition probabilities from a vertex $x$ to its neighbouring vertices are same, we say that the QW on the multilayer network is unbiased. To model such an unbiased QW, we can attach a Fourier coin $F^{(x)}$ to each vertex $x$ given by
\begin{equation}\label{Fourier coin}
    F^{(x)}=\frac{1}{\sqrt{d_x}}\sum_{r=1}^{d_x}\sum_{s=1}^{d_x}e^{2i\pi (r-1)(s-1)/d_x}|r\rangle\langle s|
\end{equation}
where $d_x$ is the degree of vertex $x$ \cite{mukai2020discrete}. Note that, the relationship of $F^{(x)}(F^{(x)})^{\dagger}=(F^{(x)})^{\dagger}F^{(x)}=\mathbb{I}$ is preserved by the Fourier coin. In the studies of QWs on graphs, it has been shown that a QW with Grover coin tends to be localized around the initial vertex \cite{mukai2020discrete}. Hence, it is worth exploring Grover walk on a multilayer network as well. The $ij^{th}$ element of the Grover coin $G^{(x)}$ attached to the vertex $x$ on a multilayer network can be written as
\begin{equation}\label{Grover coin}
    G^{(x)}_{ij}=\begin{cases}
     \frac{2-d_x}{d_x}       & \quad \text{If} \ i=j \\
     \frac{2}{d_x}  & \quad \text{Otherwise}
   \end{cases} 
\end{equation}
where $1 \leq i,j \leq d_x$ and $d_x$ is the degree of vertex $x$. Note that, Grover coin holds the relationship of $G^{(x)}(G^{(x)})^{\dagger} =(G^{(x)})^{\dagger}G^{(x)}=\mathbb{I}$. In section \ref{Analysis 1} and \ref{Analysis 2}, we analyze the dynamics of some QWs on specific multilayer networks with different choices of coin operators.
\section{Classical random walk on Multilayer networks} \label{CRW_on_MLN}
The propagation of the classical random walk (CRW) on different graph structures is a topic that has been studied extensively \cite{lovasz1993random}. In general, the propagation of a random walker on any given network structure is modeled using the transition probabilities from a vertex $x$ to its neighbouring vertices \cite{venegas2008quantum, kendon2006quantum}. By adopting the same concept, one can define a CRW on a multilayer network as well \cite{baptista2022universal}. Let $\Omega_{x,y}$ be the transition probability from vertex $x$ to $y$. Then, the probability $P_{c}(x,t)$ of finding the random walker at position $x$ at time $t$ is given by the following recurrence relations
\begin{equation}\label{CRW equation}
   P_{c}(x,t)=\sum_{r=1}^{d_x}\Omega_{f_x(r),x}P_{c}(f_x(r),t-1)
\end{equation}
where $d_x$ is the degree of vertex $x$ and for each $r$, the function $f_x(r)$ gives the labels of the neighbouring vertices of $x$. When the transition probabilities from a vertex $x$ to its neighbouring vertices are same, we say that the CRW is unbiased. In usual practice \cite{venegas2008quantum}, transition probability $\Omega^{(ub)}_{x,y}$ for an unbiased CRW on any graph structure is defined as follows
\begin{equation}\label{unbiased CRW equation}
    \Omega^{(ub)}_{x,y}=\begin{cases}
     \frac{1}{d_x} & \quad \text{If} \ (x,y)\ \text{is connected} \\
     0  & \quad \text{Otherwise}\\
   \end{cases}
\end{equation}
where $d_x$ is the degree of vertex $x$. One can adopt the same definition given in \eqref{unbiased CRW equation} to model an unbiased CRW on a multilayer network. 
\section{Numerical implementation on a toy model} \label{Analysis 1}
\begin{figure}[h]
\begin{center}
\includegraphics[width=2in, height=2in]{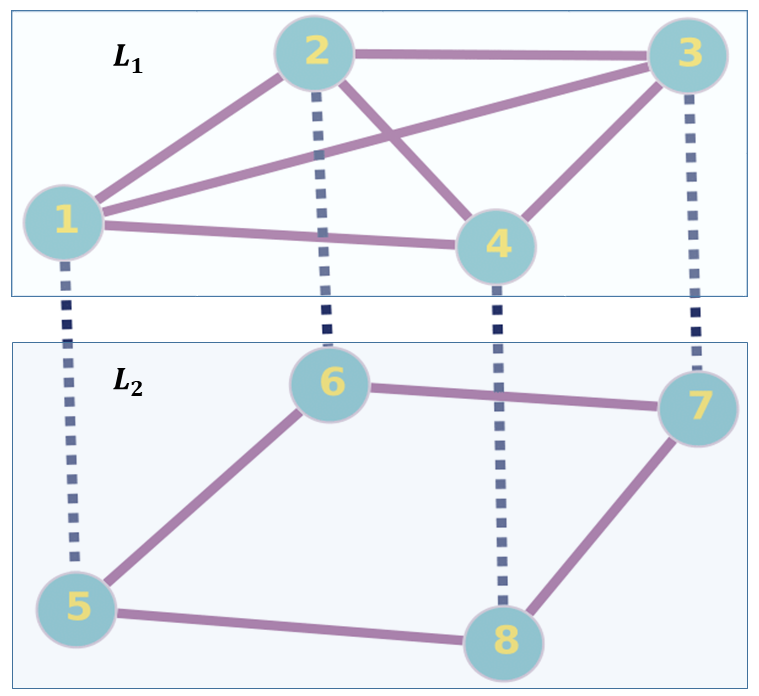}
\caption{Schematic diagram of a multilayer network with four vertices and two layers named $L_1$ and $L_2$. The set of labels $\{x,x+4\}$ where $x \in \{1,2,3,4\}$ represent the same entity in different layers. The solid lines connect the vertices within each layer and the dotted lines connect inter-layer. Since the inter-layer edges link only the vertices representing the same entity in different layers, this network can be classified as a multiplex network which is a special class of multilayer networks. Top layer represents an undirected regular graph and the bottom layer represents an undirected connected graph. Parallel edges and self-loops are excluded here. This schematic diagram is inspired by \cite{li2010genome,granell2013dynamical,granell2014competing,gomez2013diffusion}. }
\label{multilayer_graph_of_four_vertexes}
\end{center}
\end{figure}

In this section, we perform a CRW and a QW on an illustrated toy multilayer network structure (Figure \ref{multilayer_graph_of_four_vertexes}) and examine the flow of probability through various layers. Our intention is to explore the fundamental differences between classical and quantum dynamics on a mulitlayer network. From the definition of the mulitlayer network given in section \ref{QW_on_multilayer}, one can consider diverse configurations of structures with multilayers. Nonetheless, a two-layer network, which consists of two distinct graphs can be understood as the simplest multilayer network. Hence, here we consider one such simplest multilayer structure to perform a CRW and a QW. Since the inter-layer edges of the toy model in Figure \ref{multilayer_graph_of_four_vertexes} link only the vertices representing the same entity in different layers, this network can be classified as a multiplex network which is a special class of multilayer networks \cite{gomez2013diffusion}. Let the top and bottom layers be $\mathcal{L}_{1}=\{V_1,E_1\}$ and $\mathcal{L}_{2}=\{V_2,E_2\}$ respectively where $V_1=\{1, 2, 3 ,4\}$ and $V_2=\{5, 6, 7, 8\}$ are the set of vertices in each layer and $E_1$, $E_2$ are the set of edges corresponding to each layer. The supra-adjacency matrix \cite{boccaletti2014structure} of the multilayer network in Figure \ref{multilayer_graph_of_four_vertexes} can be written as follows

\begin{equation}\label{supra-adjacency matrix}
    A_{sup} =\begin{pmatrix}
 \FirstLayer
  & \rvline & \bigzero \\
\hline
  \bigzero & \rvline & \SecondLayer
\end{pmatrix}=
\begin{pmatrix}
  \begin{matrix}
  0 & 1 & 1 & 1 \\ 
 1 & 0 & 1 & 1 \\ 
 1 & 1 & 0 & 1 \\ 
 1 & 1 & 1 & 0 \\ 
  \end{matrix}
  & \rvline & 
 \begin{matrix} 
  1 & 0 & 0 & 0 \\ 
 0 & 1 & 0 & 0 \\ 
 0 & 0 & 1 & 0 \\ 
 0 & 0 & 0 & 1 \\
 \end{matrix} \\
\hline
  \begin{matrix} 
  1 & 0 & 0 & 0 \\ 
 0 & 1 & 0 & 0 \\ 
 0 & 0 & 1 & 0 \\ 
 0 & 0 & 0 & 1 \\
 \end{matrix}  & \rvline &
  \begin{matrix}
   0 & 1 & 0 & 1 \\ 
 1 & 0 & 1 & 0 \\ 
 0 & 1 & 0 & 1 \\ 
 1 & 0 & 1 & 0 \\ 
  \end{matrix}
\end{pmatrix}
\end{equation}

where the first ($A^{(1)}$) and second ($A^{(2)}$) block matrices along the diagonal (i.e. top left corner and bottom right corner) represent the adjacency matrices of the layer $\mathcal{L}_{1}$ and $\mathcal{L}_{2}$ respectively. Top right and bottom left block matrices represent the connection between the layers. Using the supra-adjacency matrix in \eqref{supra-adjacency matrix}, one can define the sets of $\{\mathcal{B}_{x}\}_{x=1}^{8}$ along with the set of functions $\{f_x(r)\}_{x=1}^{8}$ and then simulate the evolution of the QW on the multilayer network by updating the elements of the block matrix in \eqref{block_matrix_A}. 

\subsection{Probability Distribution}

In this section, we perform an unbiased QW and CRW on the toy multilayer network structure and investigate the probability of finding the walker on each layer after 100 steps. Recall that, to perform an unbiased QW, one needs to use the Fourier coin given in \eqref{Fourier coin}. For the case of unbiased CRW, one needs to use the transition probabilities given in \eqref{unbiased CRW equation}. We initialize the CRW from the vertex 1. For the QW, we consider two localized initial states of the form $|1\rangle_p \otimes |3\rangle_c$ and $|1\rangle_p \otimes |5\rangle_c$. That is, at the beginning we, place the quantum walker at the position state $|1\rangle_p$ (vertex 1) attaching the coin state of $|3\rangle_c$ and $|5\rangle_c$ to the walker. We select these localized initial states to replicate conditions similar to those in CRW, which enables a meaningful comparison between CRWs and QWs. After 100 steps, we calculate the probability of finding the walker at each layer by summing the probability of finding the walker at each node corresponding to a given layer. 

\begin{figure}[h] 
\begin{center}
     \includegraphics[width=2.5in,height=1.5in]{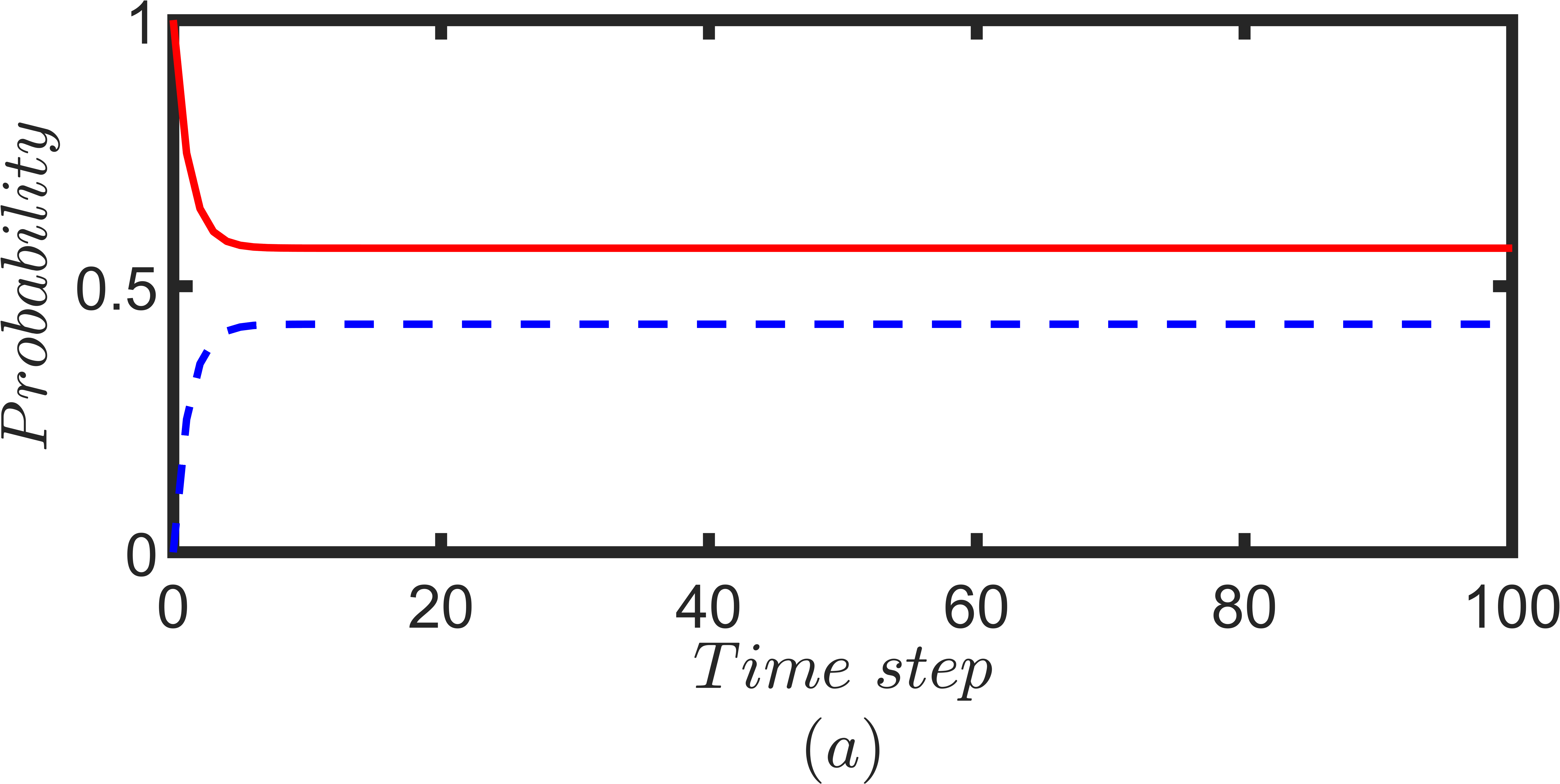} 
         \includegraphics[width=2.5in, height=1.5in]{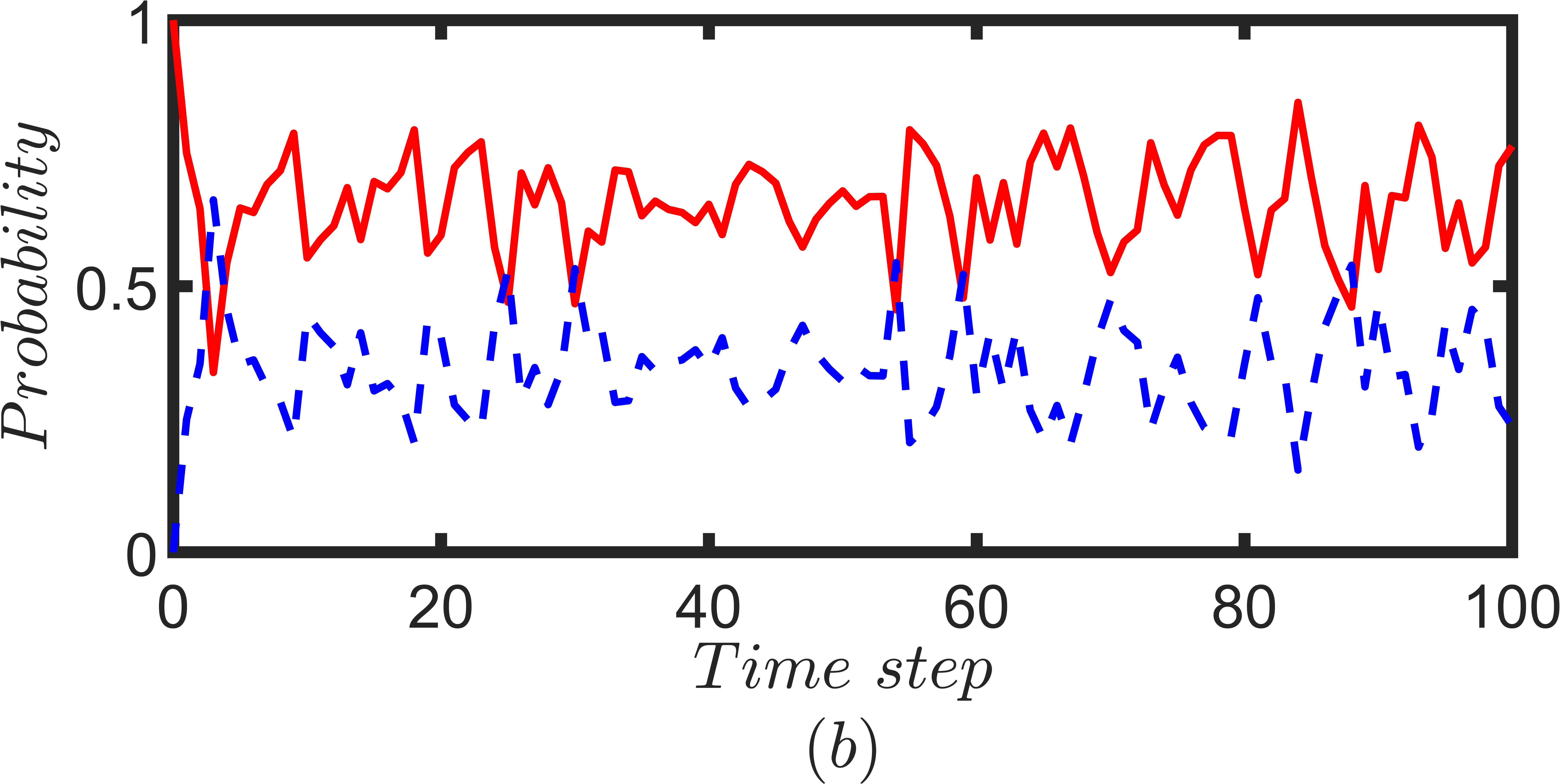} 
         \includegraphics[width=2.5in, height=1.5in]{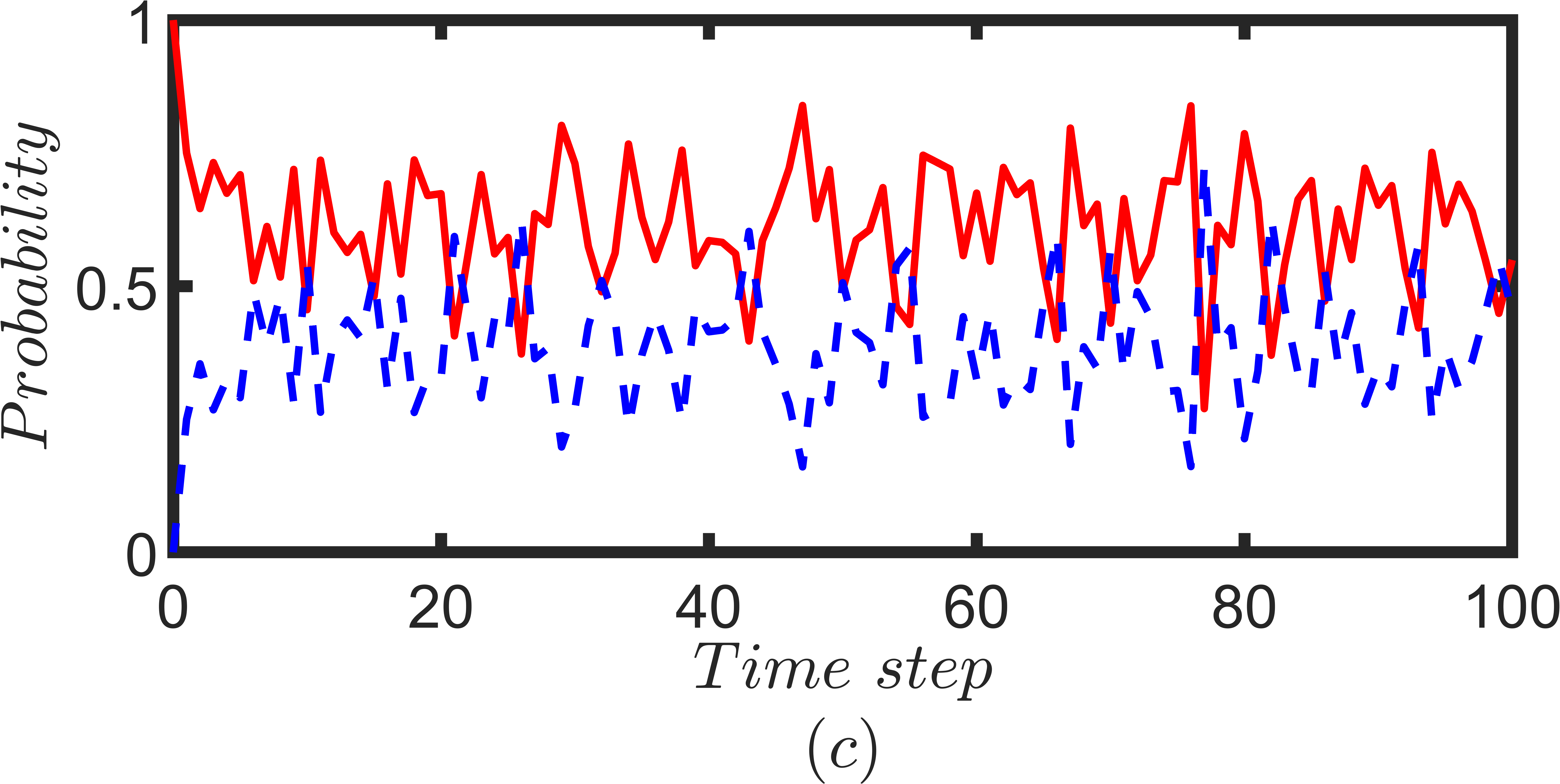}
     \caption{Probability of finding the walker on top layer (red solid line) and bottom layer (blue dotted line) for each time step up to 100 steps (a) unbiased CRW is initiated from vertex 1. Fourier walk is initiated from (b) $|1\rangle_p \otimes |3\rangle_c$ and (c) $|1\rangle_p \otimes |5\rangle_c$}.
     \label{Prob_dis_Layer1_and_2}
\end{center}
\end{figure}

An interesting distinction between the unbiased classical and quantum walkers is illustrated in Figure \ref{Prob_dis_Layer1_and_2}. For the case of CRW, the probabilities of finding the walker on the top and bottom layers eventually stabilize to a steady state as time progresses. This can be seen in Figure \ref{Prob_dis_Layer1_and_2}(a). Conversely, in QW, the quantum walker displays dynamic changes in probability over time (see Figure \ref{Prob_dis_Layer1_and_2}(b) and \ref{Prob_dis_Layer1_and_2}(c)). It is important to note that, in the case of CRWs, the probability of locating the walker on the top layer is consistently higher than that of the bottom layer, which is expected since we initialize the walk from the top layer. However, in QWs, despite the initial placement on the top layer, there exists certain time steps where the probability of locating the walker on the bottom layer surpasses that of the top layer. For instead, in Figure \ref{Prob_dis_Layer1_and_2}(c) one can see certain time steps where the probability of finding the walker on the bottom layer is higher than that of the top layer. In addition, from Figure \ref{Prob_dis_Layer1_and_2}(b) and \ref{Prob_dis_Layer1_and_2}(c), one can identify that even though the initial position state is the same, different initial coin states of the QWs can control the temporarily transition of the quantum walker from top layer to bottom layer. Such a behaviour has no analogy to CRWs. Hence, unlike a classical walker who tend to stay within the initial layer, the ability of a quantum walker to temporarily transition to the bottom layer with higher probability implies that the QW could explore a broader portion of the multilayer structure. This enhanced exploration could be useful when searching through large, complex databases represented as multilayer networks. In section \ref{Analysis 2}, we further examine this behaviour by employing different types of lager mulilayer networks.

\begin{figure}[h] 
\begin{center}
    \includegraphics[width=1.8in,height=1.4in]{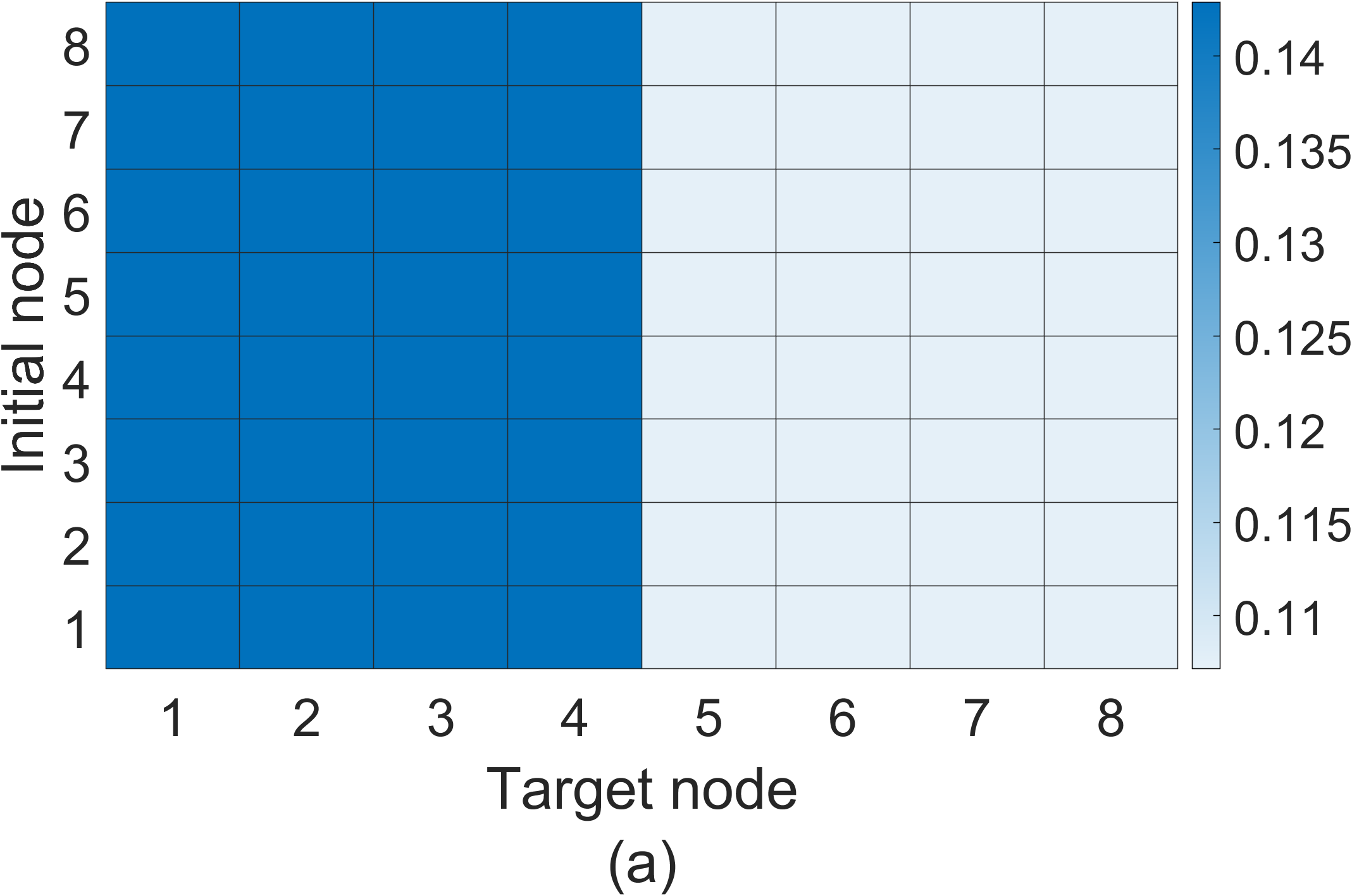} 
        \includegraphics[width=1.7in, height=1.4in]{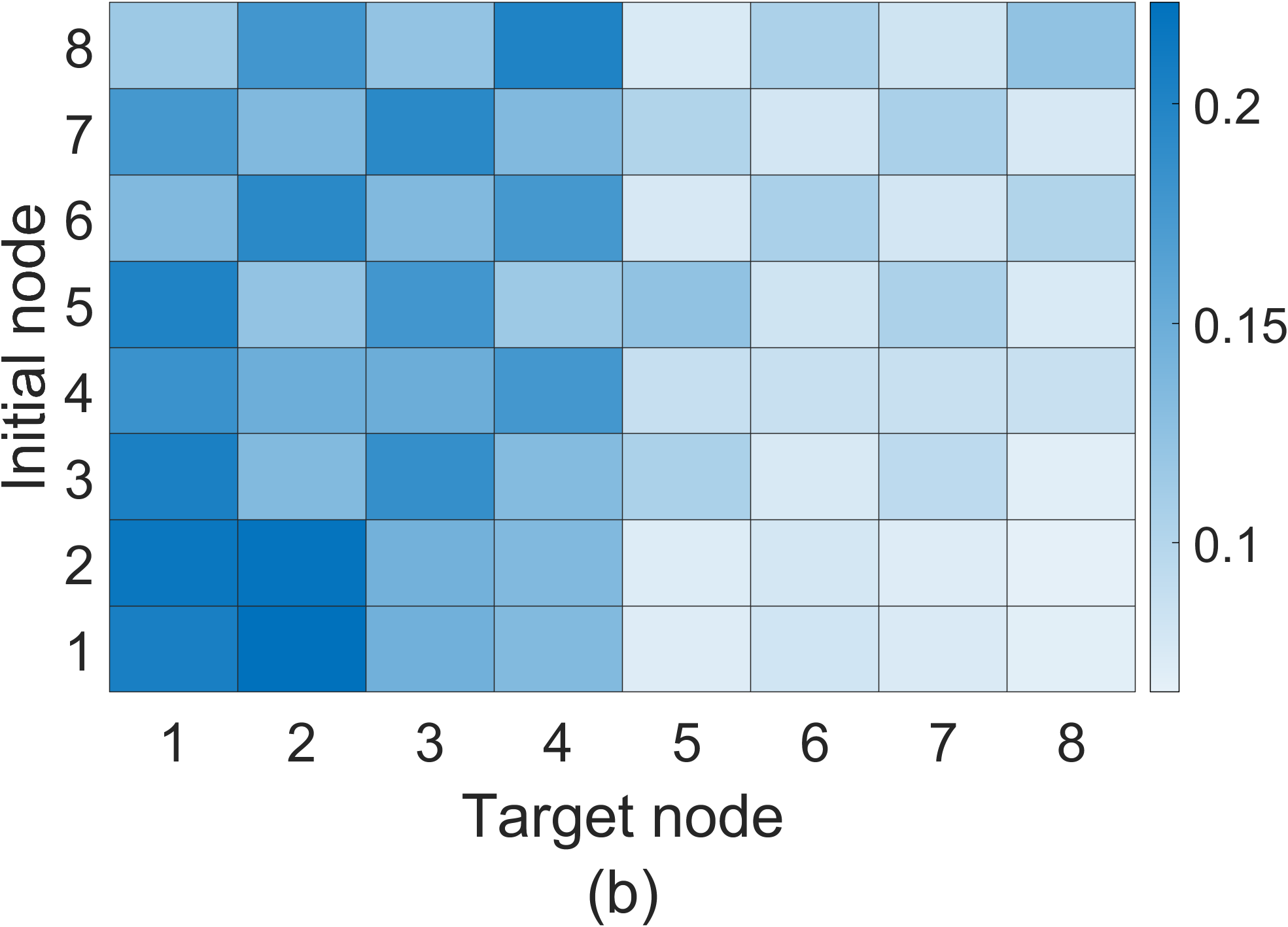} 
        \includegraphics[width=1.7in, height=1.4in]{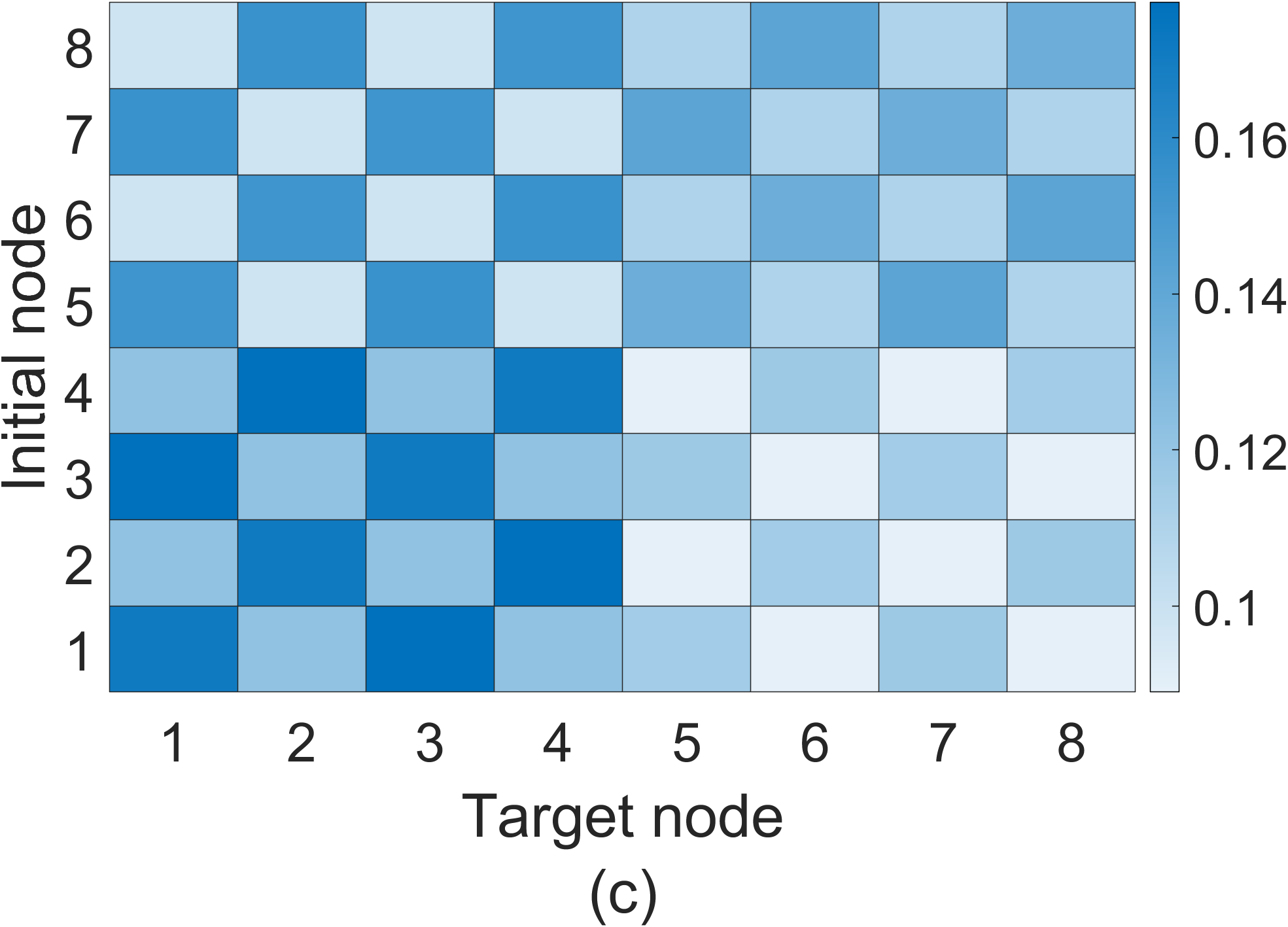} 
        \includegraphics[width=1.7in, height=1.4in]{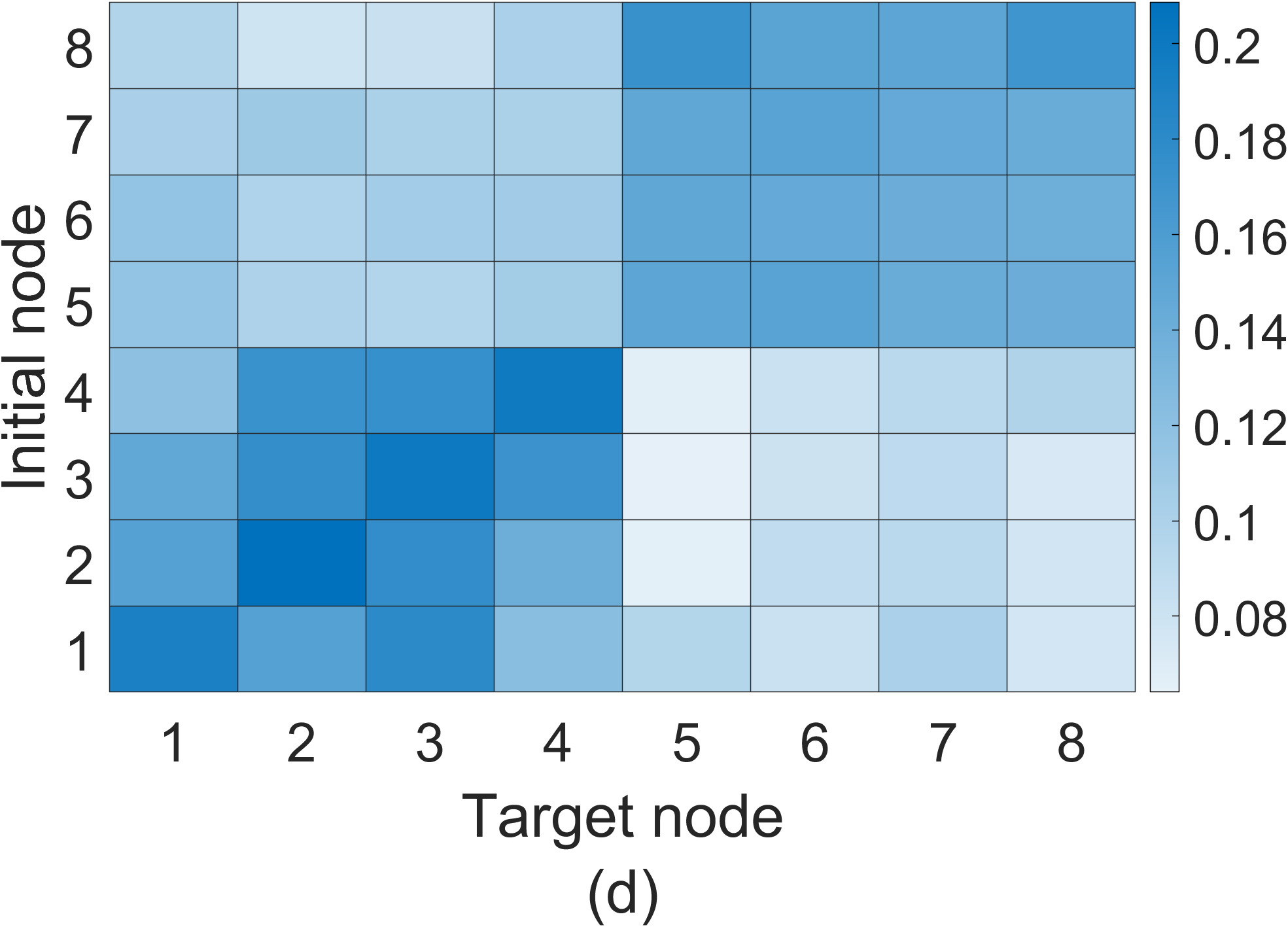}
        \includegraphics[width=1.7in, height=1.4in]{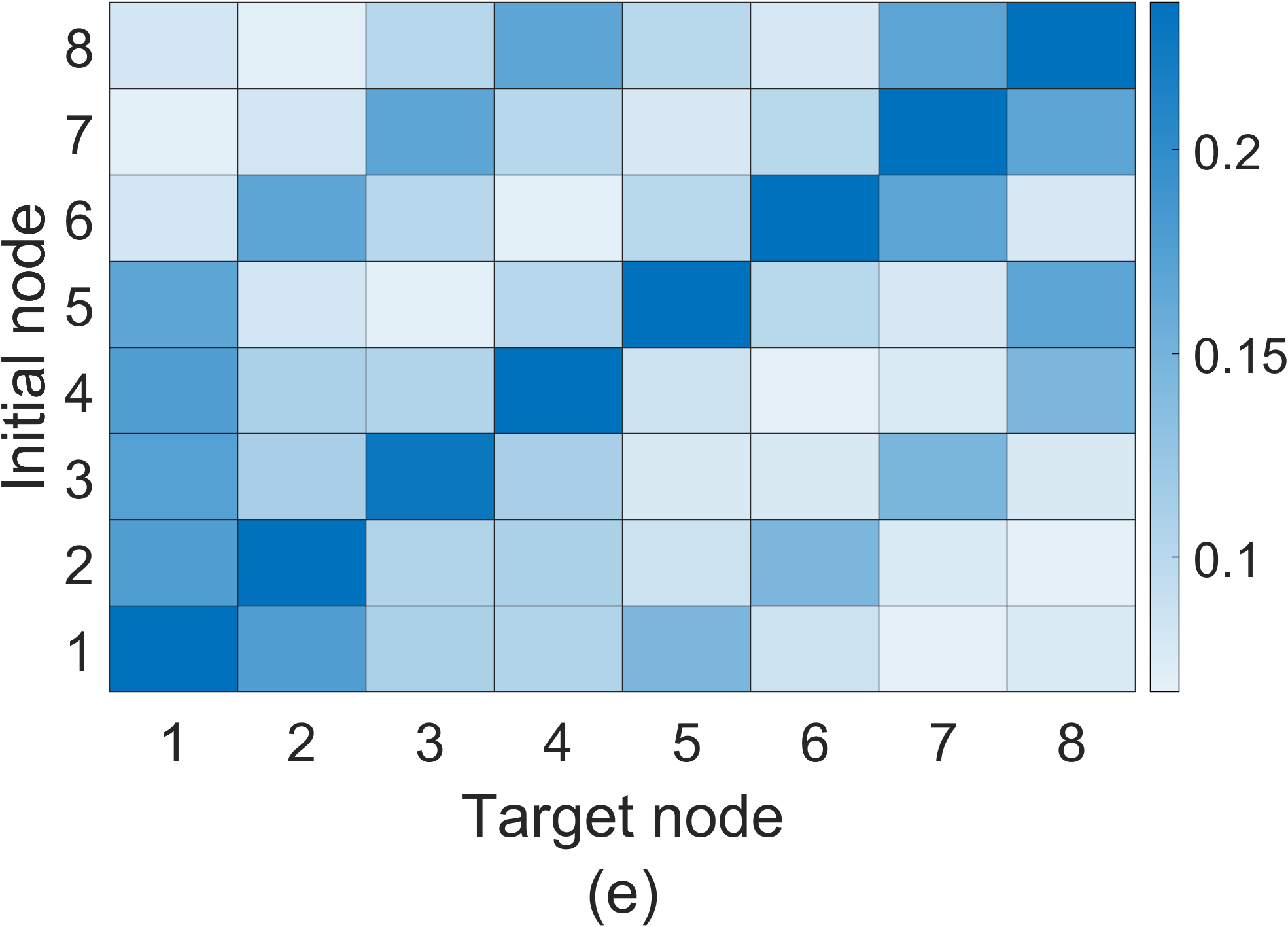}
     \caption{Heatmaps are depicted for (a) unbiased CRW, (b,d) Fourier walk and (c,e) Grover walk. The vertical axis shows the initial node from which the walker starts the walk and the horizontal axis shows the target node where the walker ends the walk. Each square corresponding to the Fourier and Grover walks indicates the value of time-averaged probabilities for a time period of $T =100$. For the unbiased CRW, each square corresponds to the probability after 100 time steps. Both Fourier and Grover walks in (b) and (c) are initiated from the localized position state of the form $|\phi_1\rangle \equiv |x\rangle_p \otimes \frac{1}{\sqrt{d_x}}\sum_{r=1}^{d_x}|f_x(r)\rangle_c$. The Fourier and Grover walks in (d) and (e) are initiated from the localized position state of the form $|\phi_2\rangle\equiv|x\rangle_p \otimes \bigg(\frac{i}{\sqrt{d_x}}|f_x(1)\rangle_c+\frac{1}{\sqrt{d_x}}\sum_{r=2}^{d_x-1}|f_x(r)\rangle_c-\frac{i}{\sqrt{d_x}}|f_x(d_x)\rangle_c\bigg)$. Note that, both localized initial states $|\phi_1\rangle$ and $|\phi_2\rangle$ have a uniform superposition of coin states, yet  $|\phi_2\rangle$ contains some complex coefficients.}
     \label{Time-avaerage}
\end{center}
\end{figure}

In Figure \ref{Prob_dis_Layer1_and_2}(b) and \ref{Prob_dis_Layer1_and_2}(c) we have seen that the quantum probability is fluctuating. It is a widely acknowledged fact that, the unitary characteristic of QWs prevents the quantum walker from achieving a steady state \cite{faccin2013degree}. Hence, to obtain an idea of the static picture, one can calculate the time-averaged probability of finding the quantum walker at vertex $x$ defined by
\begin{equation} \label{time-average probability}
    \overline{P}_{T}(x)=\frac{1}{T}\sum_{t=0}^{T-1}P_q(x,t)
\end{equation}
where $T \in \mathbb{N}$ \cite{Wang}. Note that, when $T$ becomes lager, $\overline{P}_{T}(x)$ becomes a better measure that depicts the static picture. In Figure \ref{Time-avaerage} we illustrate the time-averaged probabilities for a time period of $T = 100$ for Fourier and Grover walks. The vertical axis of the heatmaps shows the initial node from which the walker starts the walk and the horizontal axis shows the target node where the walker ends the walk. For both Fourier and Grover walk, we consider two cases. In the first case, we initiate both Fourier and Grover walker from the localized initial state of the form $|\phi_1\rangle \equiv|x\rangle_p \otimes \frac{1}{\sqrt{d_x}}\sum_{r=1}^{d_x}|f_x(r)\rangle_c$ where $x \in \{1,...,8\}$. That is, we start both QWs from each node by attaching a uniform superposition of coin states. Then, for each node, the time-averaged probability for a time period of $T = 100$ is calculated using \eqref{time-average probability}. The results are given in Figure \ref{Time-avaerage}(b) and \ref{Time-avaerage}(c). In the second case, we repeat the same procedure by using the localized initial state of the form $|\phi_2\rangle \equiv|x\rangle_p \otimes \bigg(\frac{i}{\sqrt{d_x}}|f_x(1)\rangle_c+\frac{1}{\sqrt{d_x}}\sum_{r=2}^{d_x-1}|f_x(r)\rangle_c-\frac{i}{\sqrt{d_x}}|f_x(d_x)\rangle_c\bigg)$ where $x \in \{1,...,8\}$. The results are given in Figure \ref{Time-avaerage}(d) and \ref{Time-avaerage}(e). For the unbiased CRW, we initiate the walker from each node and calculate the probability of finding the walker at each node after 100 time steps. The result is given in Figure \ref{Time-avaerage}(a). 

According to the Figure \ref{Time-avaerage}, the classical walker tends to stay on the top layer of the toy model after 100 time steps irrespective of its initial node. Since the top layer comprises a complete graph, we can expect this result. Time-average probability profile of the Fourier walk also exhibits a similar behavior like the classical walker when initiated from $|\phi_1\rangle$. That is, the time-average probability of finding the Fourier walker on the top layer is relatively higher than that of the bottom layer irrespective of the initial node. However, when the Fourier walker is initiated from $|\phi_2\rangle$, we can confine the walk to a particular layer with a higher probability, which is visible in Figure \ref{Time-avaerage}(d). When initiated from $|\phi_1\rangle$, Grover walk exhibits no significant behaviour, but for the initial condition of $|\phi_2\rangle$ Grover walker tends to stay, with higher probability, at the initial position and at the corresponding position on the other layer. As a result, one can see a sharp line along the anti-diagonal of the grid in Figure \ref{Time-avaerage}(e) and two lines parallel to this sharp anti-diagonal line. Hence, one can control the quantum dynamics on the multilayer network by changing the initial state, which has no direct analogy in CRW.   

\subsection{Return Probability} 
\begin{figure}[h] 
\begin{center}
\includegraphics[width=3in,height=1.8in]{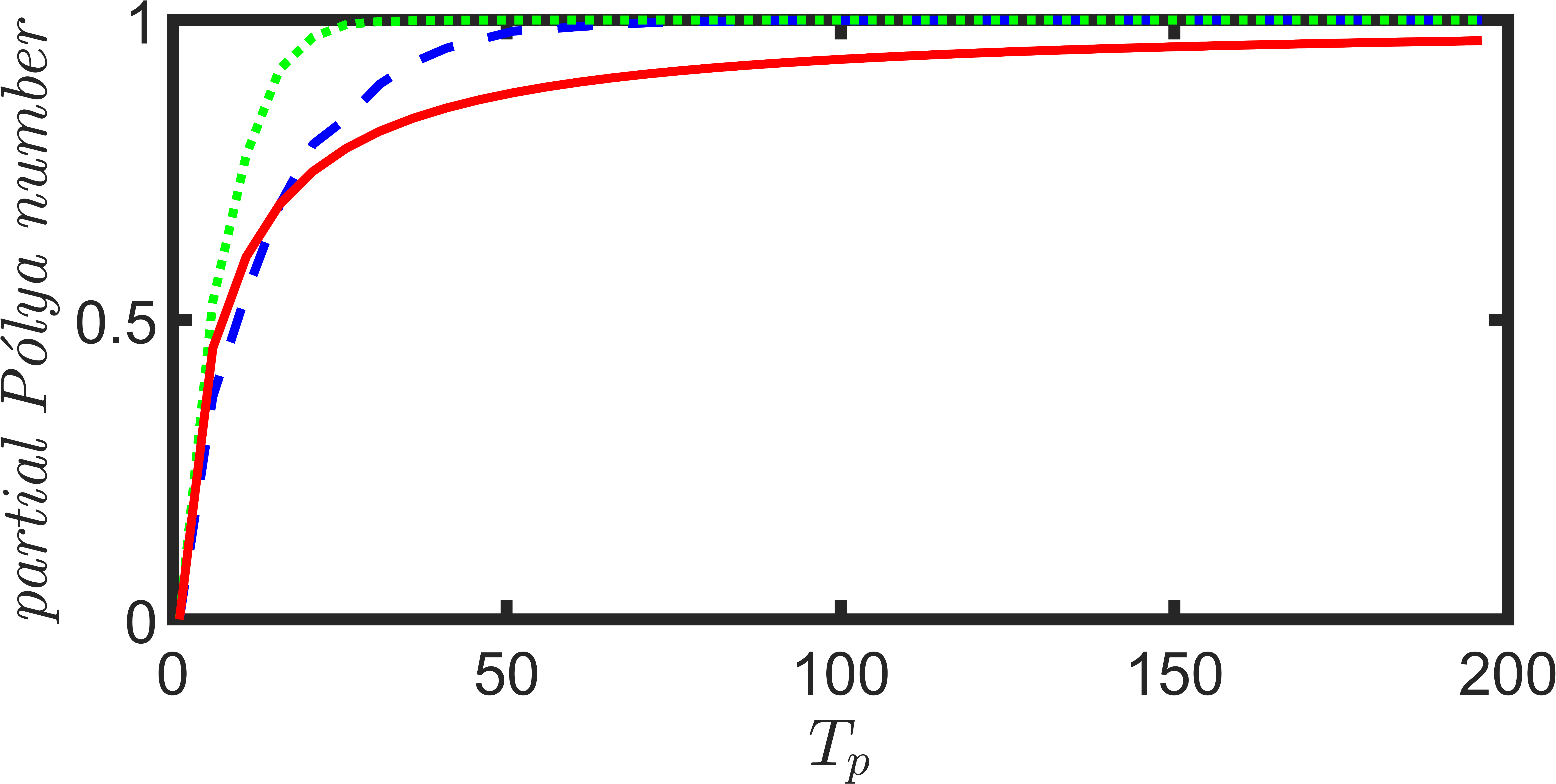} 
     \caption{Convergence of the partial P\'olya number for Grover (green dotted line), Fourier (blue dash line) and Classical (red solid line) walkers on the toy multilayer network. Partial P\'olya number is calculated by choosing a set of finite time steps $T_p \in \{1,5,10,...,200\}$ with a gap of $5$ units. Each walk is initiated from the node 1 and for both Grover and Fourier walks, initial coin state is chosen as the uniform superposition of coin states.}
     \label{recurrence}
\end{center}
\end{figure}

Another interesting question one could ask related to CRWs or QWs on a multilayer network is, how long it would take for the walker to return to it's initial position. This could be understood as the recurrence of the walk on the multilayer network. Recurrence in a CRW is characterized by the Pólya number \cite{polya1921aufgabe}, which can be written as
\begin{equation}
    P = 1 - \frac{1}{\sum_{t=1}^\infty P(x_0,t)}
    \label{eq:pnumber}
\end{equation}
where $P(x_0,t)$ is the probability of finding the walker at the initial node $x_0$ at time step $t$. For the condition of $P=1$, the walk is identified as \textit{recurrent}. Otherwise, the walk is called \textit{transient}. Moreover, the expression
\begin{equation}
    P = 1 - \Pi_{t=1}^\infty [1-P(x_0,t)]
    \label{eq:def-pnumber}
\end{equation}
can also be used as a definition for the P\'olya number as it provides the same criteria for the classification of the walk \cite{vstefavnak2008recurrencea}. With the formula given in \eqref{eq:def-pnumber}, notion of P\'olya number can be extended to the study of recurrence in QWs \cite{vstefavnak2008recurrence, vstefavnak2008recurrencea}. For practical purposes, one can calculate the partial P\'olya number using either \eqref{eq:pnumber} or \eqref{eq:def-pnumber} for a finite number of time steps $t=T_p$ rater than extending $t \rightarrow \infty$. We calculate the partial P\'olya number for CRW, Grover and Fourier walks on the toy multilayer network by choosing a set of finite time steps $T_p \in \{1,5,10,...,200\}$ with a gap of $5$ units. Our purpose is to make an estimation of the convergence of the P'olya number for each walk. We initialize the walker from node 1 and for both Grover and Fourier walks, initial coin state is chosen as the uniform superposition of coin states. (That is, $|1\rangle_p \otimes \frac{1}{\sqrt{d_1}}\sum_{r=1}^{d_1}|f_1(r)\rangle_c$). Figure \ref{recurrence} shows the convergence of the partial P\'olya number for CRW, Grover and Fourier walks.  

According to Figure \ref{recurrence}, Grover and Fourier walkers exhibit recurrence during the progression of the walk on the toy multilayer structure. However, Grover walker returns to it’s initial position faster than both Fourier and Classical walker. 

\subsection{Impact of decoherence}\label{decoherence}
In this section we study the impact of decoherence on the quantum dynamics of the walker that propagates on the toy mulitlayer network model. With regards to this, we study the impact of decoherence that arises from randomly broken links. In the context of QWs on graphs, broken link decoherence specifically relates to the loss of coherence in a QW due to the imperfections or disruptions in the graph structure \cite{leung2010coined,kollar2012asymptotic}. This can occur when edges in the graph are altered or removed, creating discontinuities in the QW. These disruptions can be caused by various factors, including physical imperfections, noise, or intentional modifications to the graph. We perform a Fourier walk for 100 time steps on the toy mulitlayer network model while breaking the connection between node 1 and 3 with a 0.5 probability at each time step. Afterwards, the mean probability distribution was calculated by averaging over 1000 trials. Figure \ref{decoherence}(c) displays the mean probability distribution of the Fourier walk over 100 time steps when subjected to the broken link decoherence model. Additionally, for comparison purposes, the probability distributions of the unbiased CRW and the standard Fourier walk are also presented in Figure \ref{decoherence}(a) and \ref{decoherence}(c) respectively.

\begin{figure}[h] 
\begin{center}
\includegraphics[width=1.7in,height=1.5in]{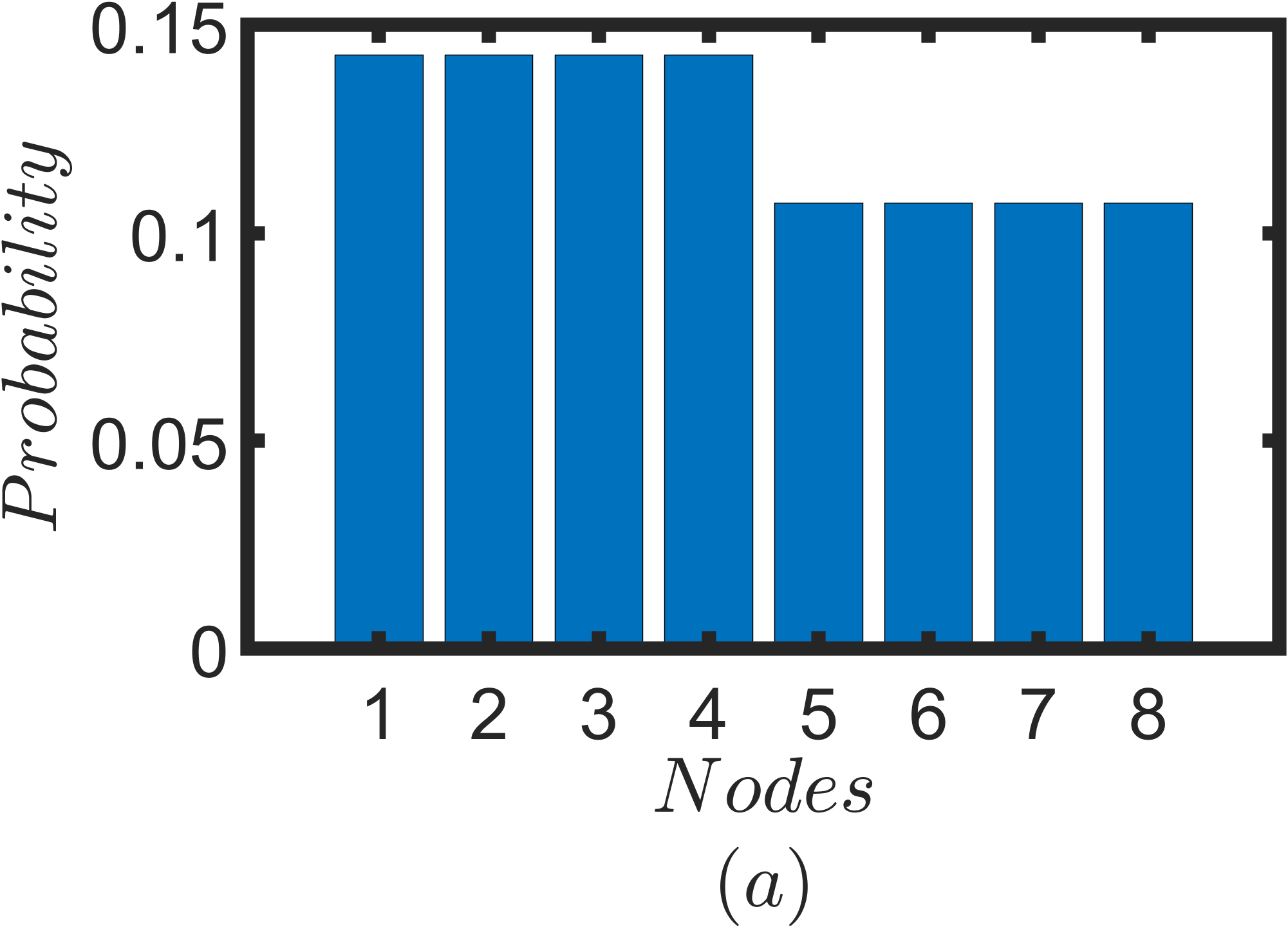}
\includegraphics[width=1.7in,height=1.5in]{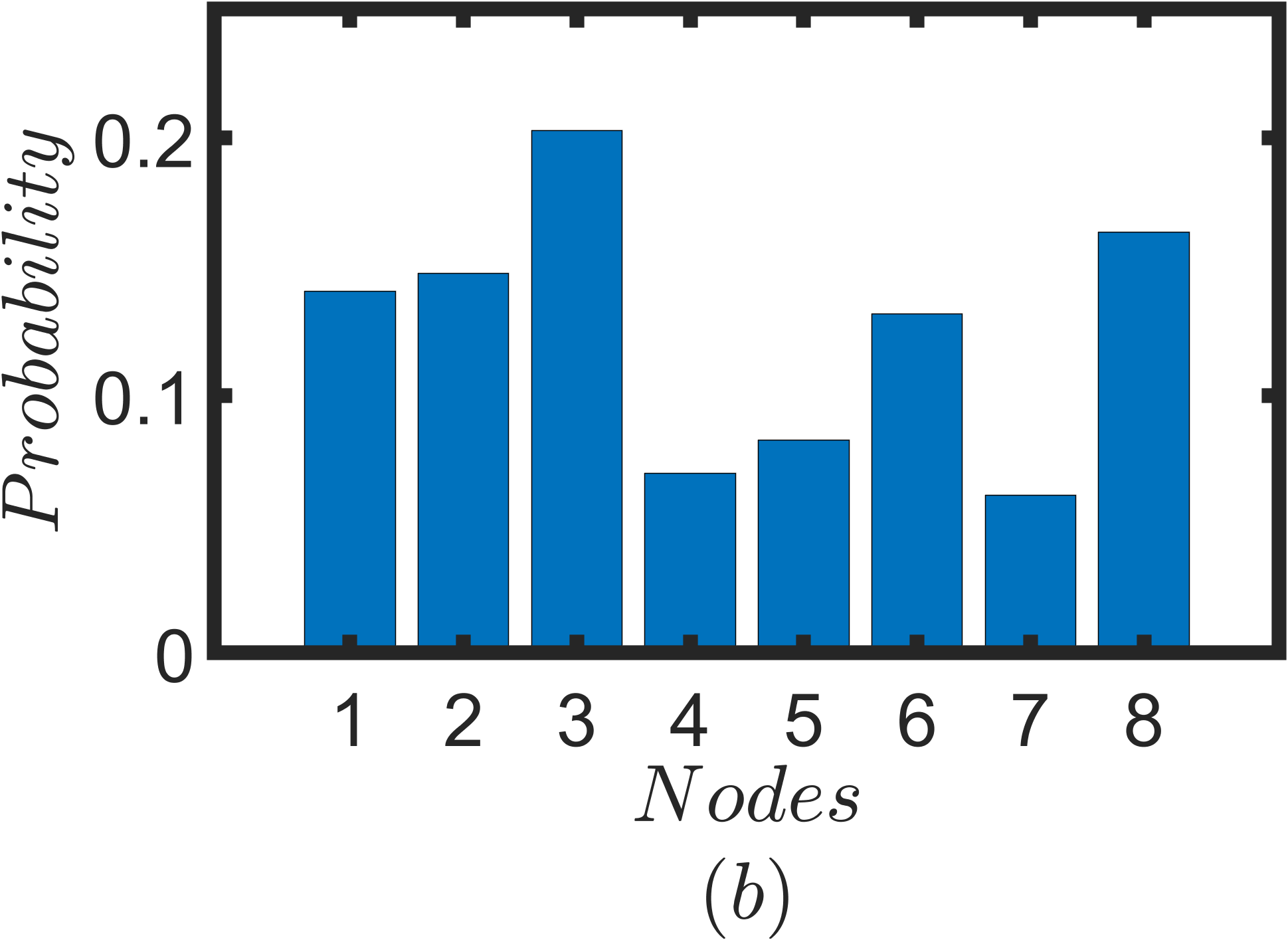}
\includegraphics[width=1.7in,height=1.5in]{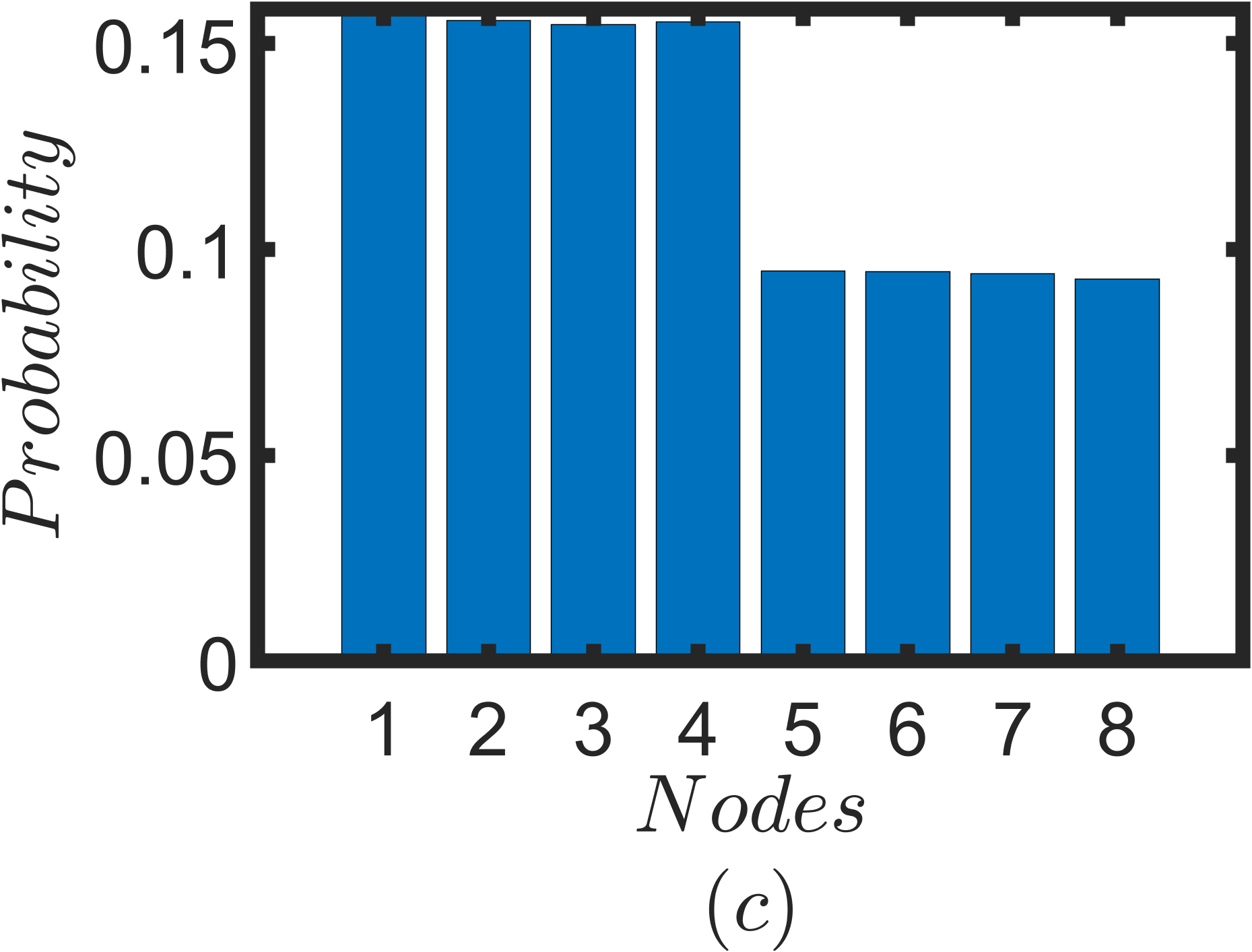}
     \caption{Probability distributions of the (a) unbiased CRW and (b) the standard Fourier walk after 100 time steps on the toy multilayer network. The CRW is initiated from node 1 and Fourier walk is initiated from the localized state of $|1\rangle_p \otimes |2\rangle_c$. The mean probability distribution of the Fourier walk subjected to broken link decoherence model is given in (c). The mean probability distribution is generated by breaking the connection between the node 1 and 3 of the toy multilayer network model with a 0.5 probability at each time step of the Fourier walk and by averaging over 1000 trials.} 
     \label{decoherence}
\end{center}
\end{figure}

According to the Figure \ref{decoherence}(a), the classical walker tends to stay on the top layer with higher probability after 100 time steps, a behaviour which we have consistently seen in Figure \ref{Prob_dis_Layer1_and_2} and \ref{Time-avaerage}. On the other hand, the standard Fourier walk exhibits a very different probability profile compered to a CRW as shown in Figure \ref{decoherence}(b). However, when subjected to broken link decoherence model, the classical signature emerges in the average probability distribution of the Fourier walk as depicted in Figure \ref{decoherence}(c). Recall that, in this study, we have realized the broken link decoherence model by breaking the connection between the node 1 and 3 of the toy multilayer network model with a 0.5 probability at each time step. That is, we are allowing to alter only a single edge in the toy multilayer network. Yet, the affect of decoherence make a substantial impact on the dynamics of walk eventually converging it to the classical distribution. This implies that QWs on a multilayer network could be very sensitive to decoherence models like broken links. 

\section{Numerical implementation on synthetic multilayer networks} \label{Analysis 2}
\begin{figure}[h] 
\begin{center}
    \includegraphics[width=2.5in,height=1.5in]{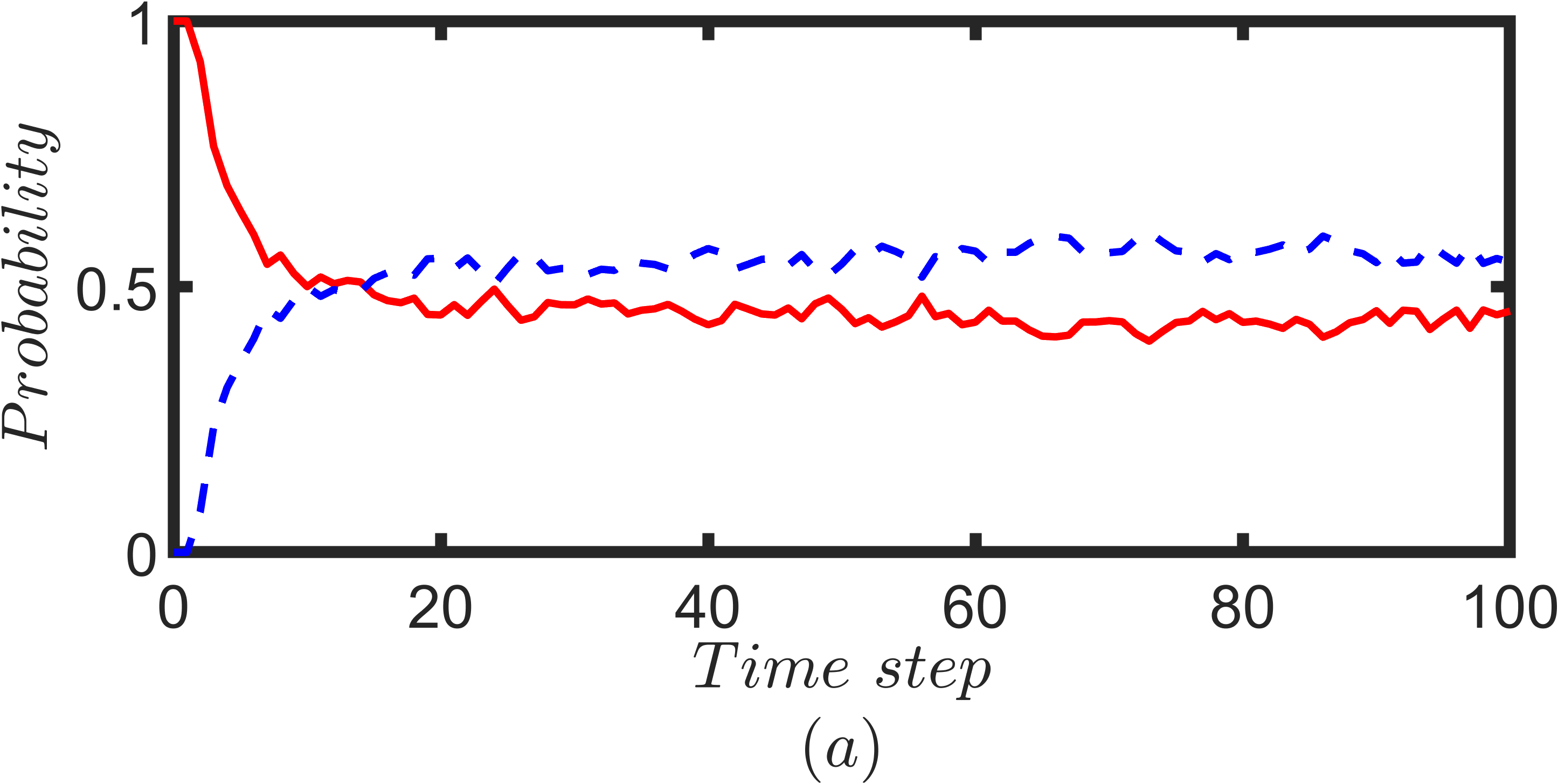} 
        \includegraphics[width=2.5in, height=1.5in]{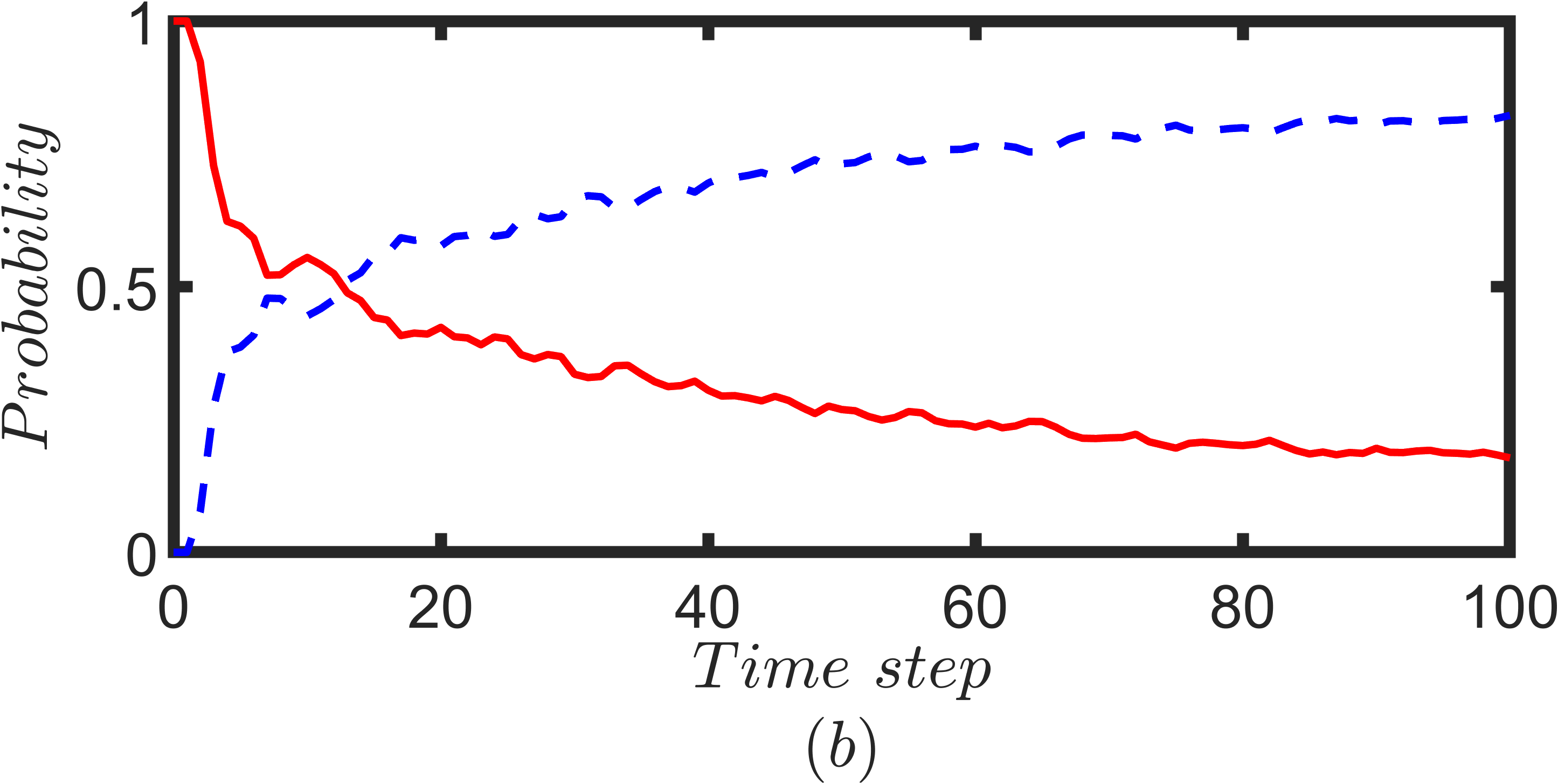} 
        \includegraphics[width=2.5in, height=1.5in]{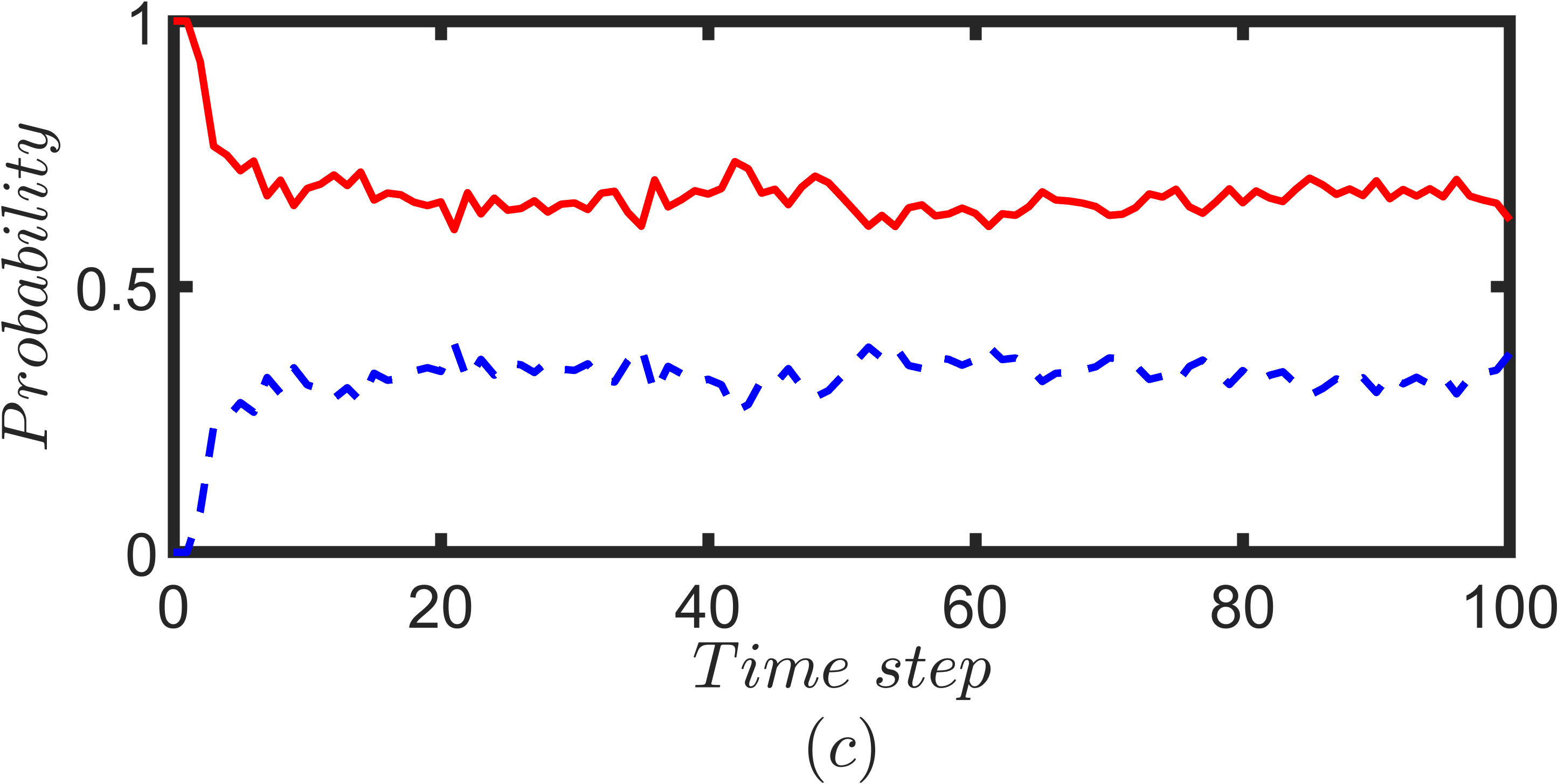} 
        \includegraphics[width=2.5in, height=1.5in]{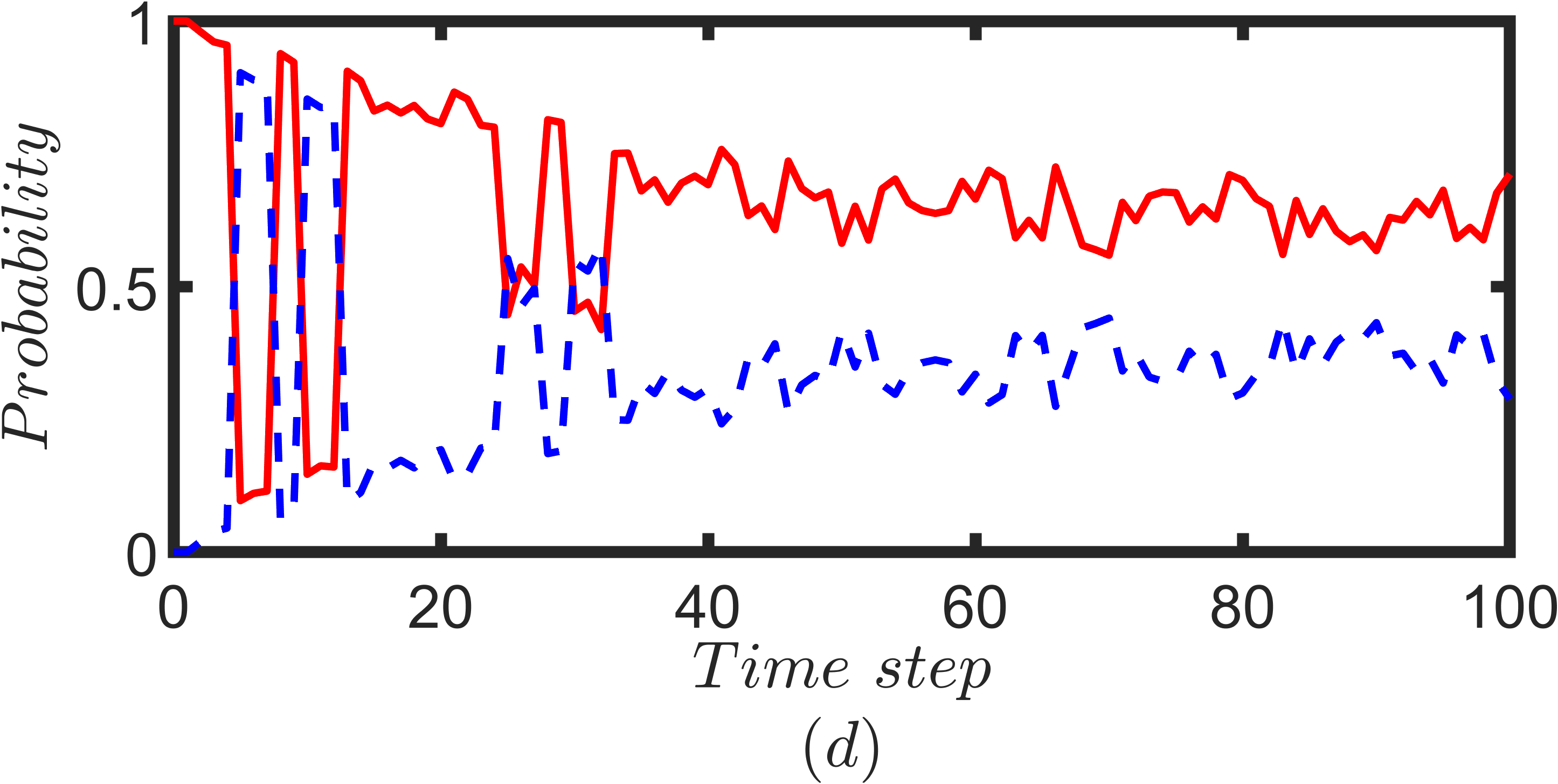}
        \includegraphics[width=2.5in, height=1.5in]{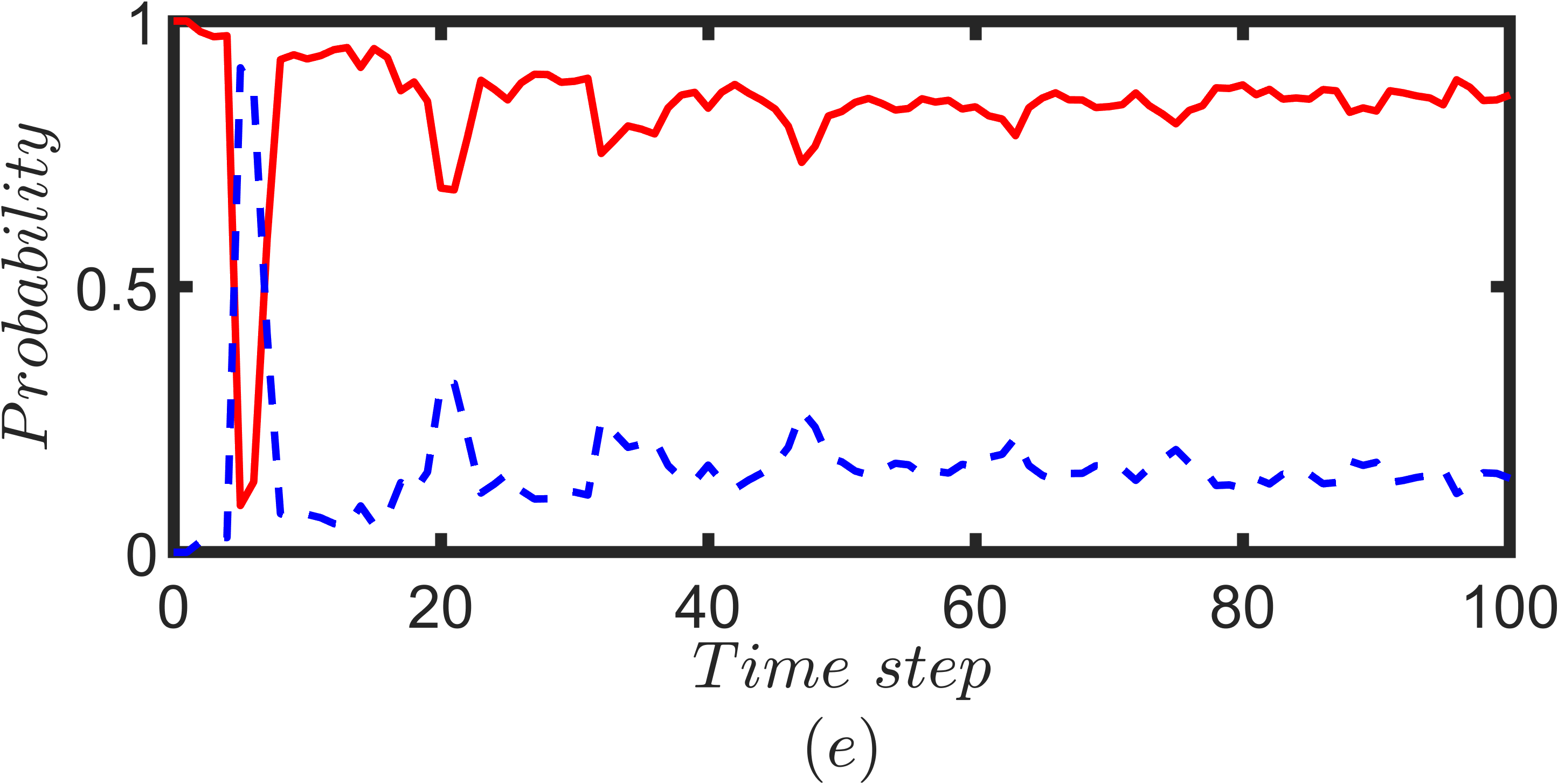}
        \includegraphics[width=2.5in, height=1.5in]{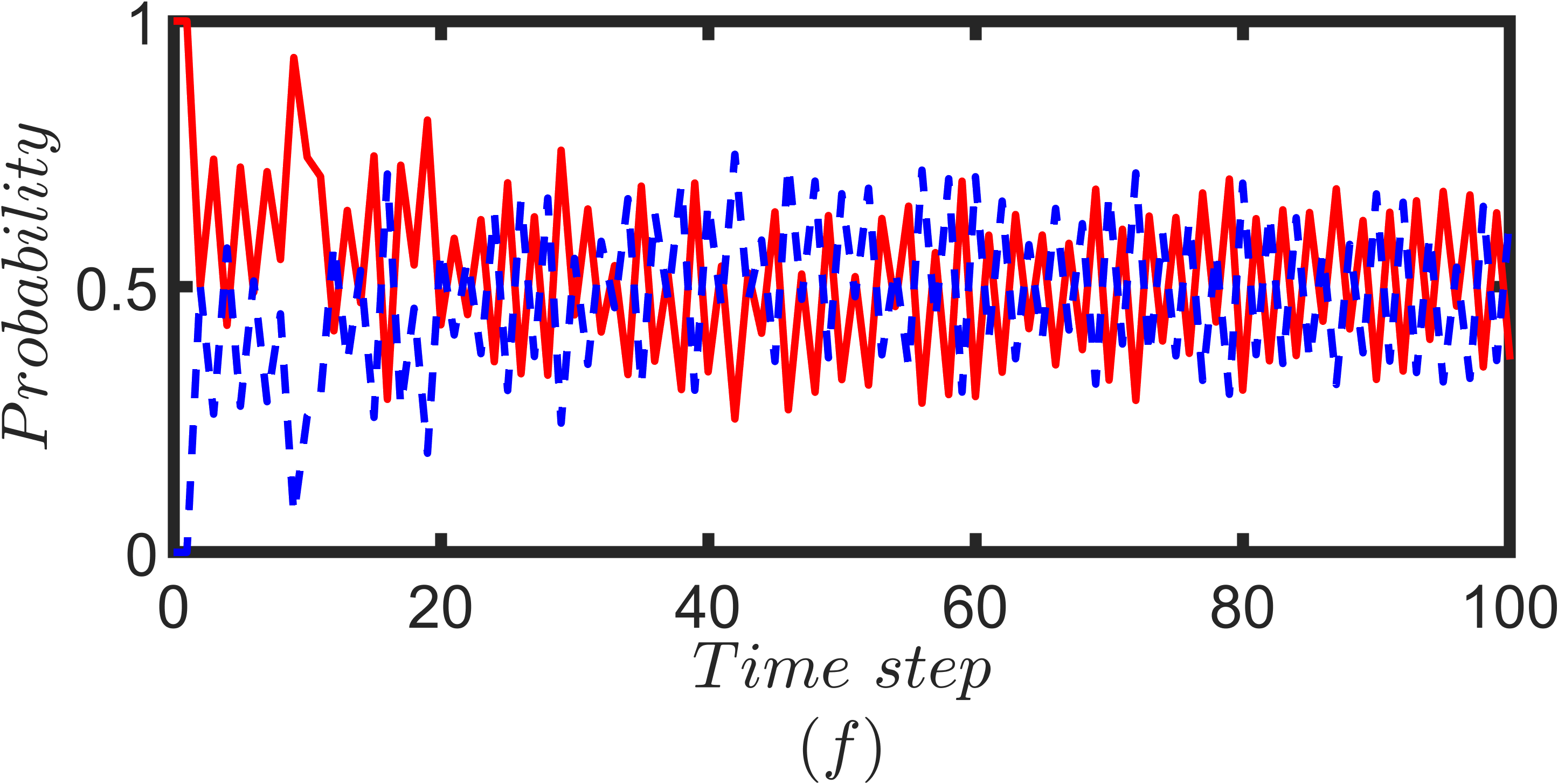}
     \caption{This figure illustrates the probability of finding the Fourier walker on the top layer (red solid line) and the bottom layer (blue dotted line) of six different two-layered multiplex networks, each consists of 100 nodes. The top and bottom layers of each multiplex network are constructed from combinations of scale-free (SF), complete (CP) and star networks with 50 nodes. (a) SF-SF (b) SF-CP (c) SF-STAR (d) CP-CP (e) CP-STAR and (f) STAR-STAR. For the case of SF-SF, two different scale-free networks are chosen. Moreover, the hub node of the star network is taken as the first node. Fourier walk is initiated from the localized position state of the form $|1\rangle_p \otimes \frac{1}{\sqrt{d_1}}\sum_{r=1}^{d_1}|f_1(r)\rangle_c$ and for each time step up to 100 steps, the probability of finding the walker on a given layer is calculated by summing the probabilities of finding the walker at each node corresponding to that layer.}
     \label{Prob_on_layers_Fourier_big_network}
\end{center}
\end{figure}

We next apply our model to perform QWs on six different two-layered multiplex networks, each consists of 100 nodes. The top and bottom layers of each multiplex network are constructed from combinations of scale-free (SF), complete (CP) and star networks with 50 nodes. For the case of SF-SF multiplex network, two different scale-free networks are chosen for top and bottom layers. In addition, the hub node of the star network is taken as the first node. First we perform a Fourier walk on each of the six different two-layered multiplex networks. The walk is initiated from the localized state of $|1\rangle_p \otimes \frac{1}{\sqrt{d_1}}\sum_{r=1}^{d_1}|f_1(r)\rangle_c$. For each time step up to 100 time steps, we calculate the probability of finding the walker on each layer by summing the probabilities of finding the walker at each node corresponding to that layer and the results are given in Figure \ref{Prob_on_layers_Fourier_big_network}. For the comparison purposes, we perform an unbiased CRW on the same two-layered multiplex network structures. The classical walker is initiated from node 1. The results are given in Figure \ref{Prob_on_layers_CRW_big_network}. By comparing the results given in Figures \ref{Prob_on_layers_Fourier_big_network} and \ref{Prob_on_layers_CRW_big_network}, one could conclude that the probability of finding the Fourier and classical walkers on most of the multiplex networks we study here follow a similarly trend. For instead, probability profiles for the combinations of SF-SF, SF-CP and SF-STAR given in Figures \ref{Prob_on_layers_Fourier_big_network}(a) \& \ref{Prob_on_layers_CRW_big_network} (a), \ref{Prob_on_layers_Fourier_big_network}(b) \& \ref{Prob_on_layers_CRW_big_network} (b) and \ref{Prob_on_layers_Fourier_big_network}(c) \& \ref{Prob_on_layers_CRW_big_network} (c) exhibit a similar trend. However, for the cases of CP-CP, CP-STAR and STAR-STAR, Fourier walk shows slight differences in its probability profiles in the first few time steps compared to the CRW. Nonetheless, as time elapses, overall trend of the probability profiles of the Fourier walker becomes similar to that of CRW. 

\begin{figure}[h] 
\begin{center}
    \includegraphics[width=2.4in,height=1.3in]{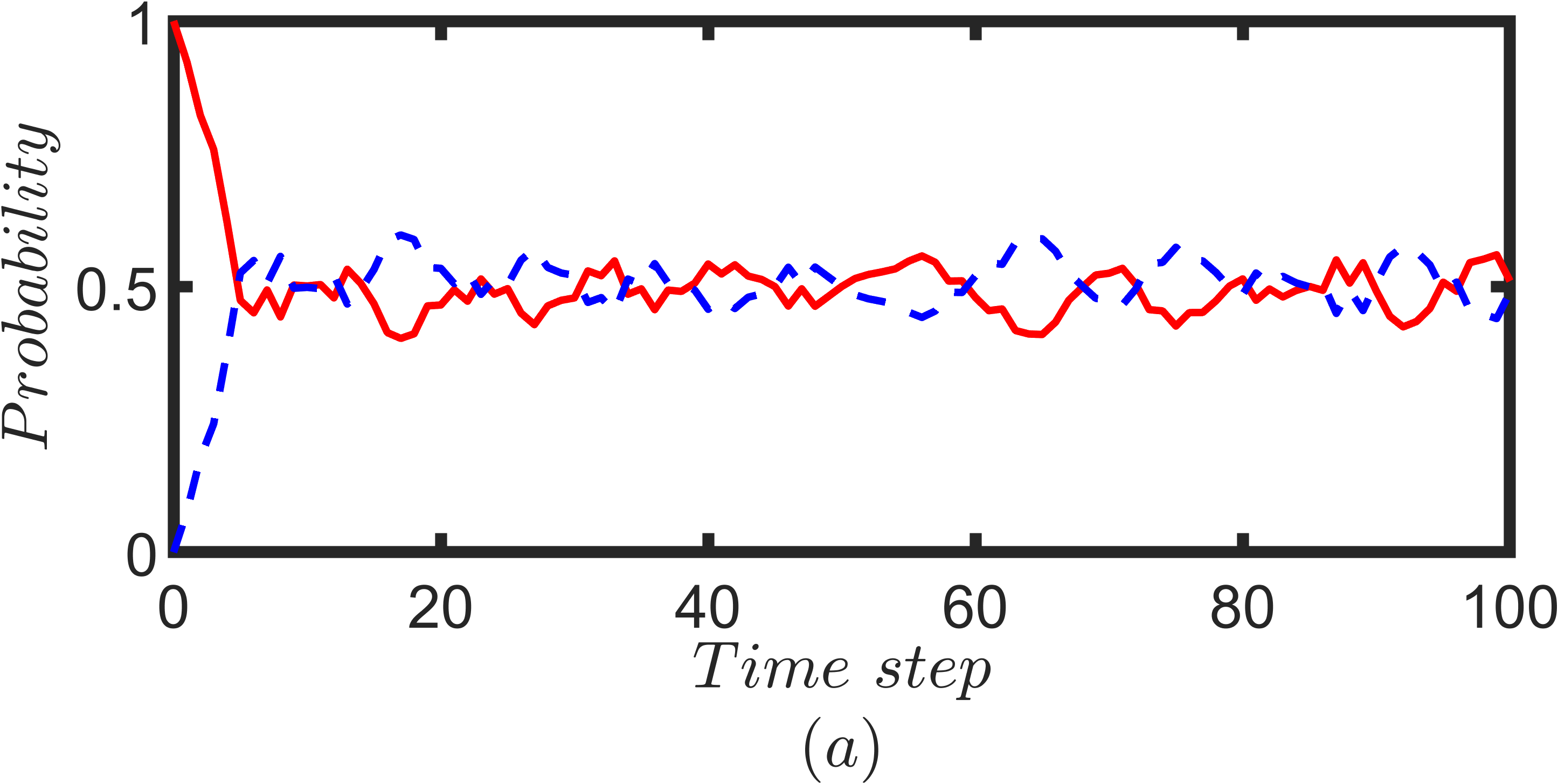} 
        \includegraphics[width=2.4in, height=1.3in]{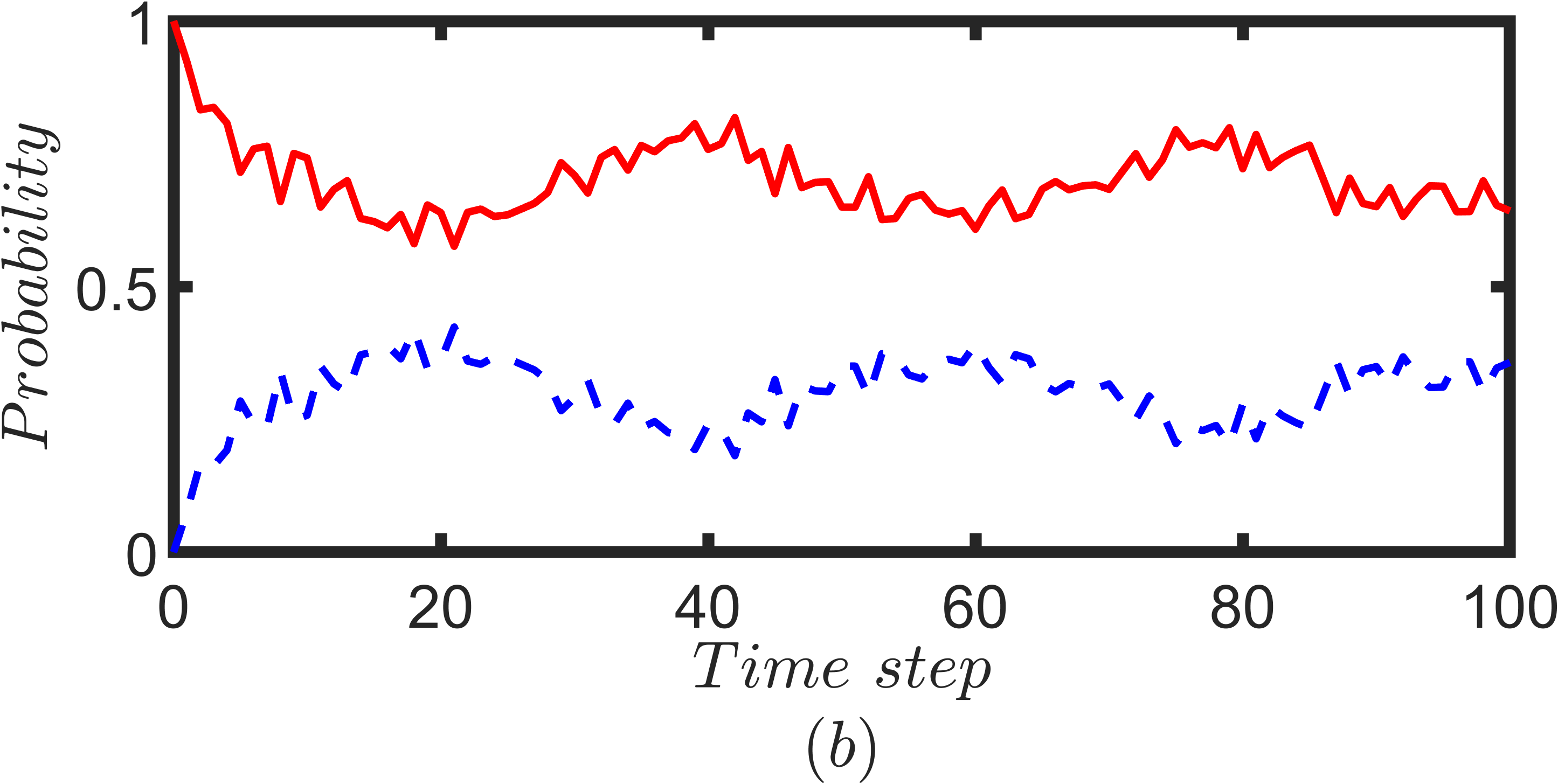} 
        \includegraphics[width=2.4in, height=1.3in]{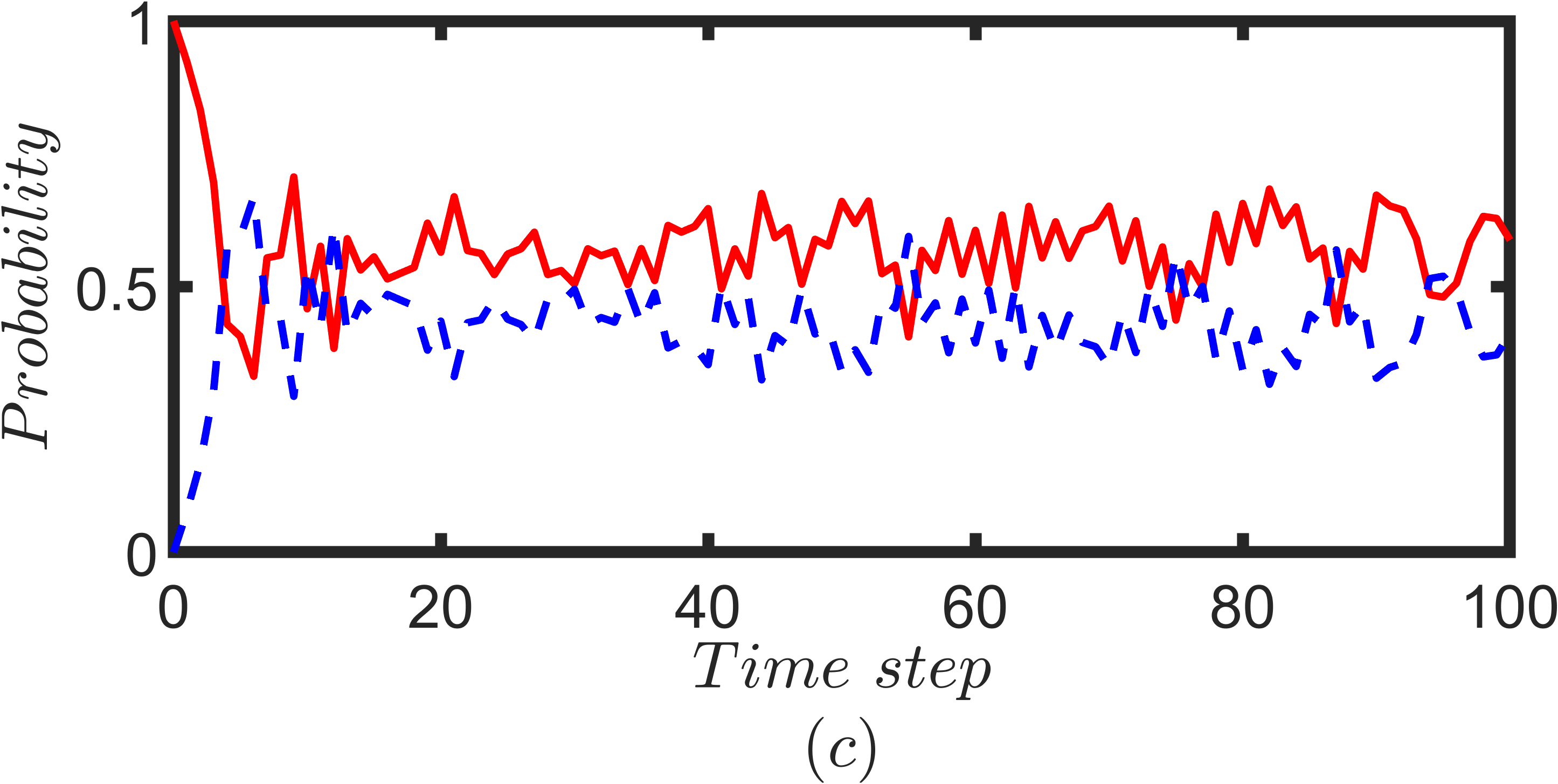} 
        \includegraphics[width=2.4in, height=1.3in]{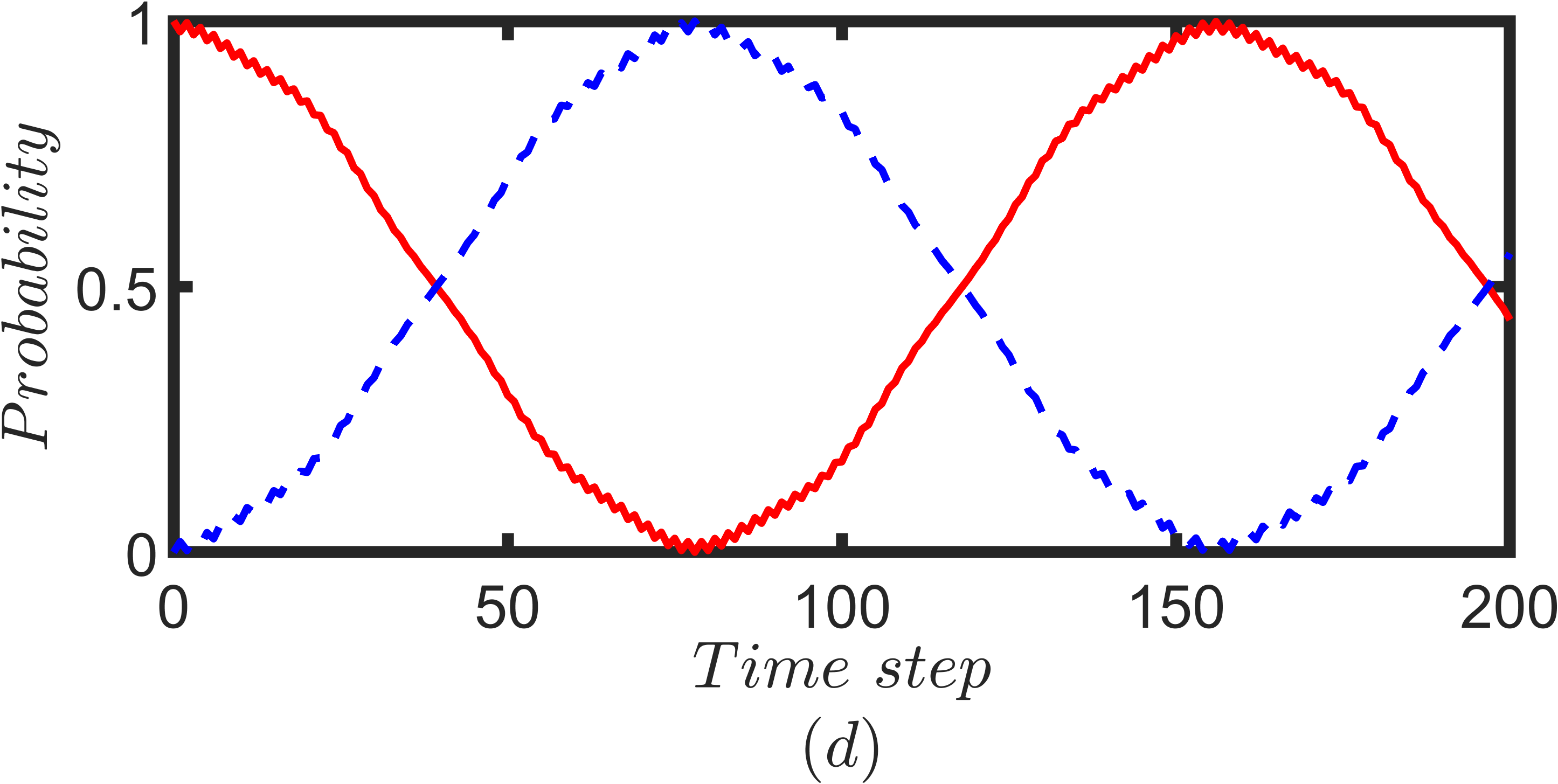}
        \includegraphics[width=2.4in, height=1.3in]{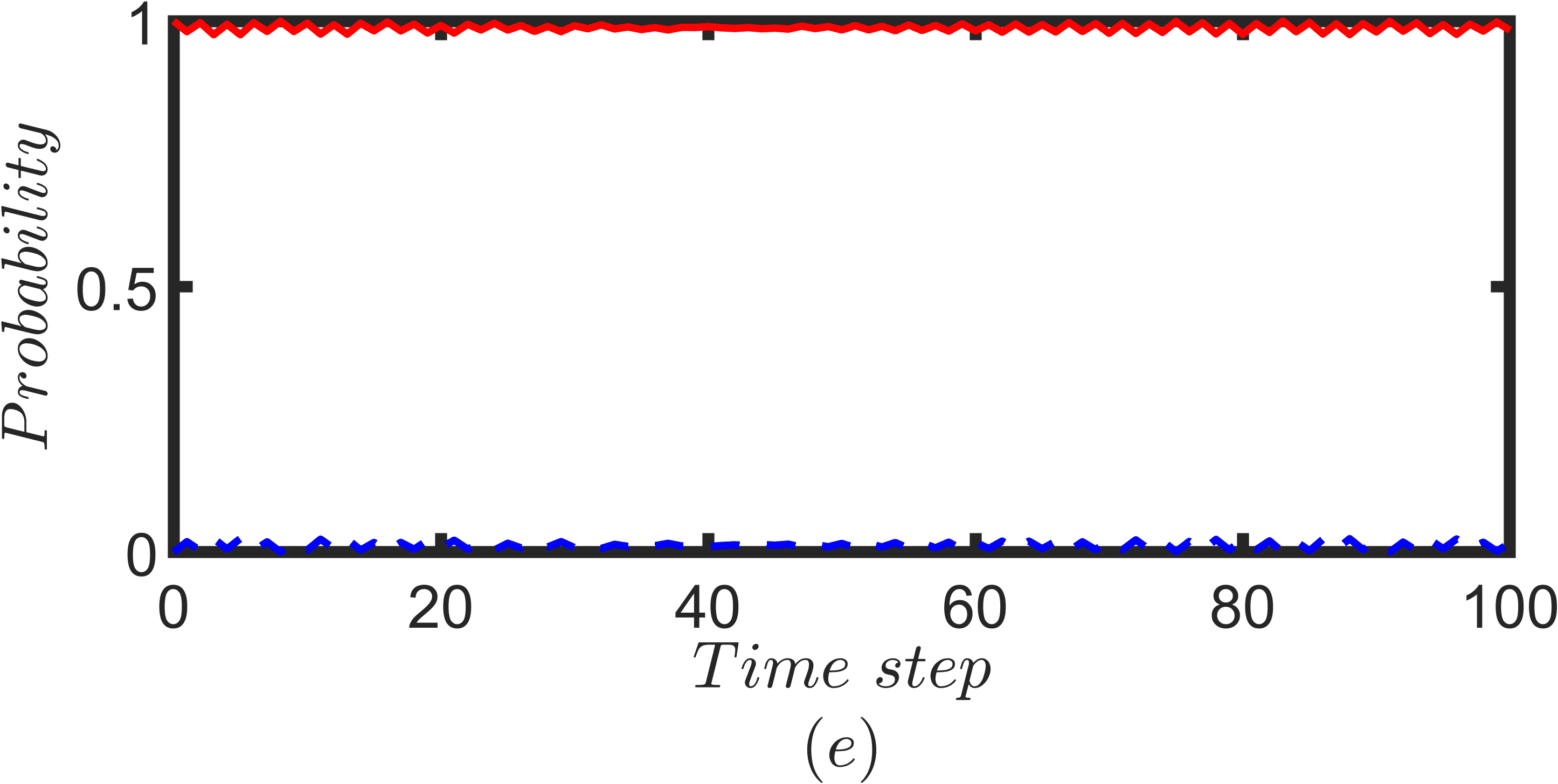}
        \includegraphics[width=2.4in, height=1.3in]{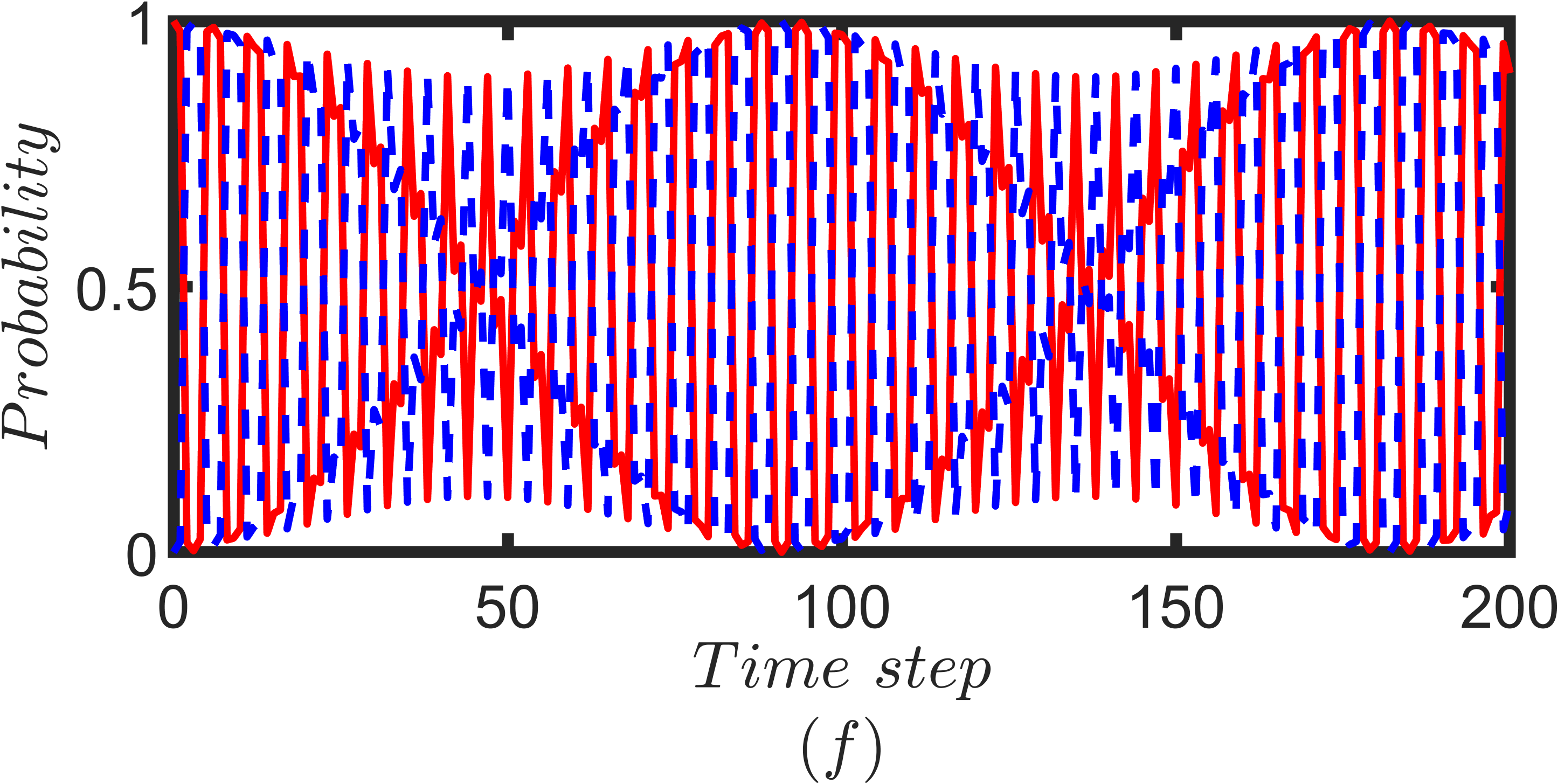}
     \caption{This figure illustrates the probability of finding the Grover walker on the top layer (red solid line) and the bottom layer (blue dotted line) of six different two-layered multiplex networks, each consists of 100 nodes. The top and bottom layers of each multiplex network are constructed from combinations of scale-free (SF), complete (CP) and star networks with 50 nodes. (a) SF-SF (b) SF-CP (c) SF-STAR (d) CP-CP (e) CP-STAR and (f) STAR-STAR. For the case of SF-SF, two different scale-free networks are chosen. Moreover, the hub node of the star network is taken as the first node. Grover walk is initiated from the localized position state of the form $|1\rangle_p \otimes \frac{1}{\sqrt{d_1}}\sum_{r=1}^{d_1}|f_1(r)\rangle_c$ and for each time step up to 100 steps, the probability of finding the walker on a given layer is calculated by summing the probabilities of finding the walker at each node corresponding to that layer. In cases (d) and (f), the time step has been extended to 200 steps to improve the clarity and visibility of the plot shapes.}
     \label{Prob_on_layers_Grover_big_network}
\end{center}
\end{figure}

The probability profiles of the Grover walk on the six different multiplex networks are given in Figure \ref{Prob_on_layers_Grover_big_network}. The Grover walker is initiated from the localized state of $|1\rangle_p \otimes \frac{1}{\sqrt{d_1}}\sum_{r=1}^{d_1}|f_1(r)\rangle_c$ and for each time step up to 100 time steps, we calculate the probability of finding the walker on each layer by summing the probabilities of finding the walker at each node corresponding to that layer. On average, for the cases of SF-SF and SF-STAR, the Grover walker tends to be on both layers with an equal probability. This can be identified from Figures \ref{Prob_on_layers_Grover_big_network} (a) and \ref{Prob_on_layers_Grover_big_network}(c). Grover walker on the CP-STAR multiplex network behaves in a very similar way like a classical walker (see Figure \ref{Prob_on_layers_Grover_big_network} (e) and \ref{Prob_on_layers_CRW_big_network} (e)). A periodic behaviour of the probability of finding the Grover walker on top and bottom layers can be seen on CP-CP and STAR-STAR multiplex networks (see Figure \ref{Prob_on_layers_Grover_big_network} (d) and \ref{Prob_on_layers_Grover_big_network}(f)). The periodicity reflects the Grover walker's coherent oscillations between the network layers and could be influenced by the topology and connectivity of the CP-CP and STAR-STAR multiplex network. Further research and analysis are essential to unlock the full potential of these insights for practical quantum applications.

\begin{figure}[h] 
\begin{center}
\includegraphics[width=2.4in,height=1.3in]{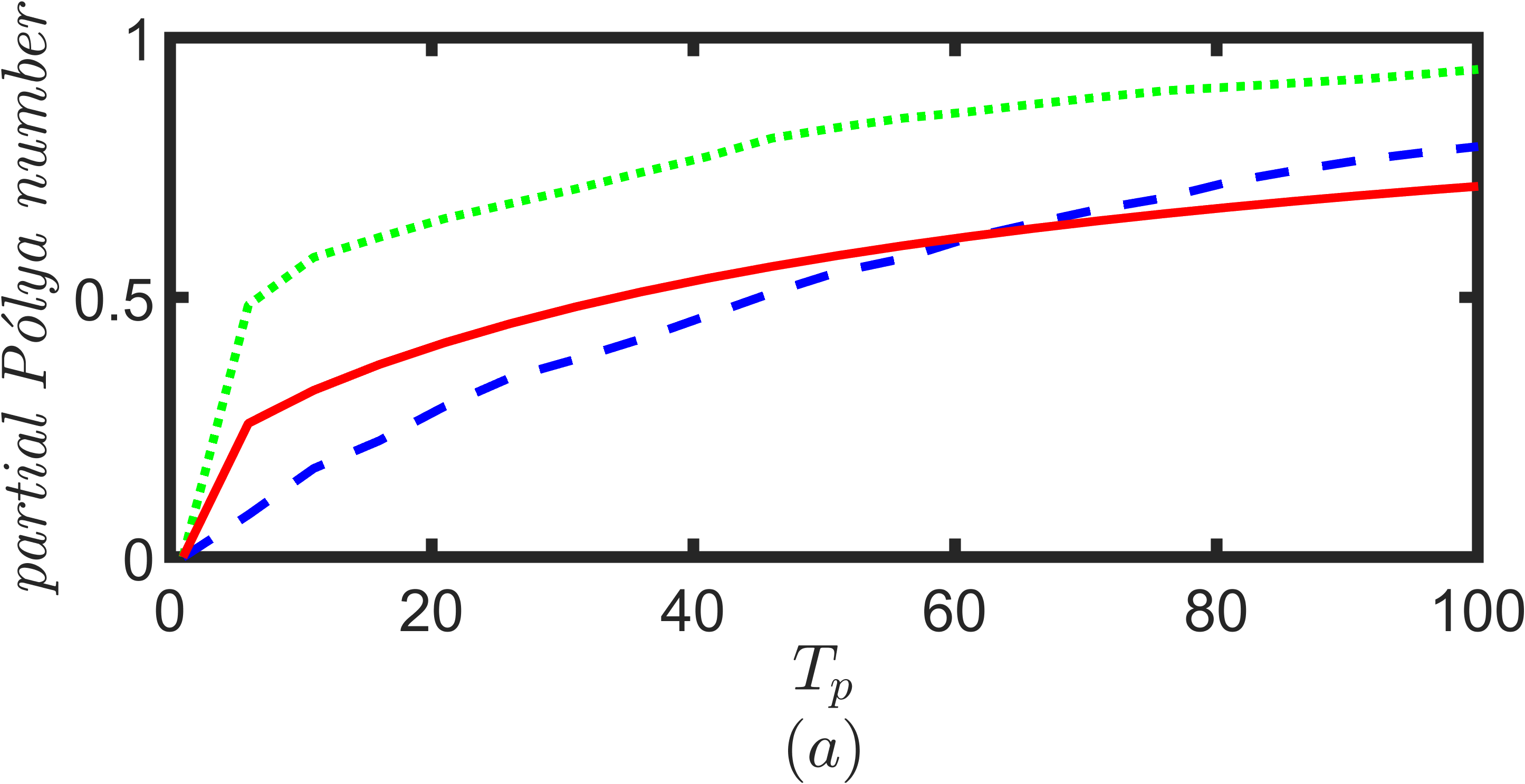}
\includegraphics[width=2.4in,height=1.3in]{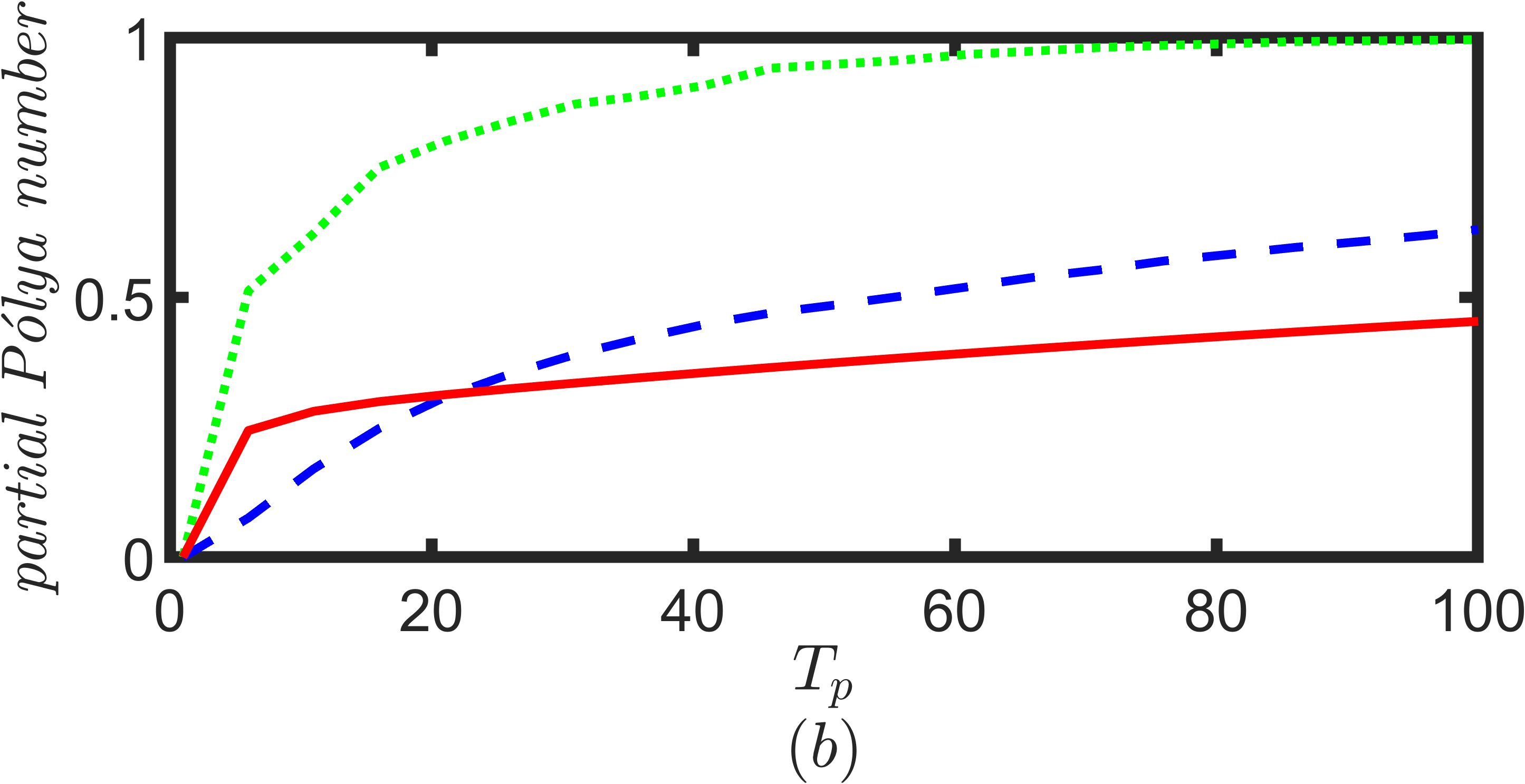} 
\includegraphics[width=2.4in,height=1.3in]{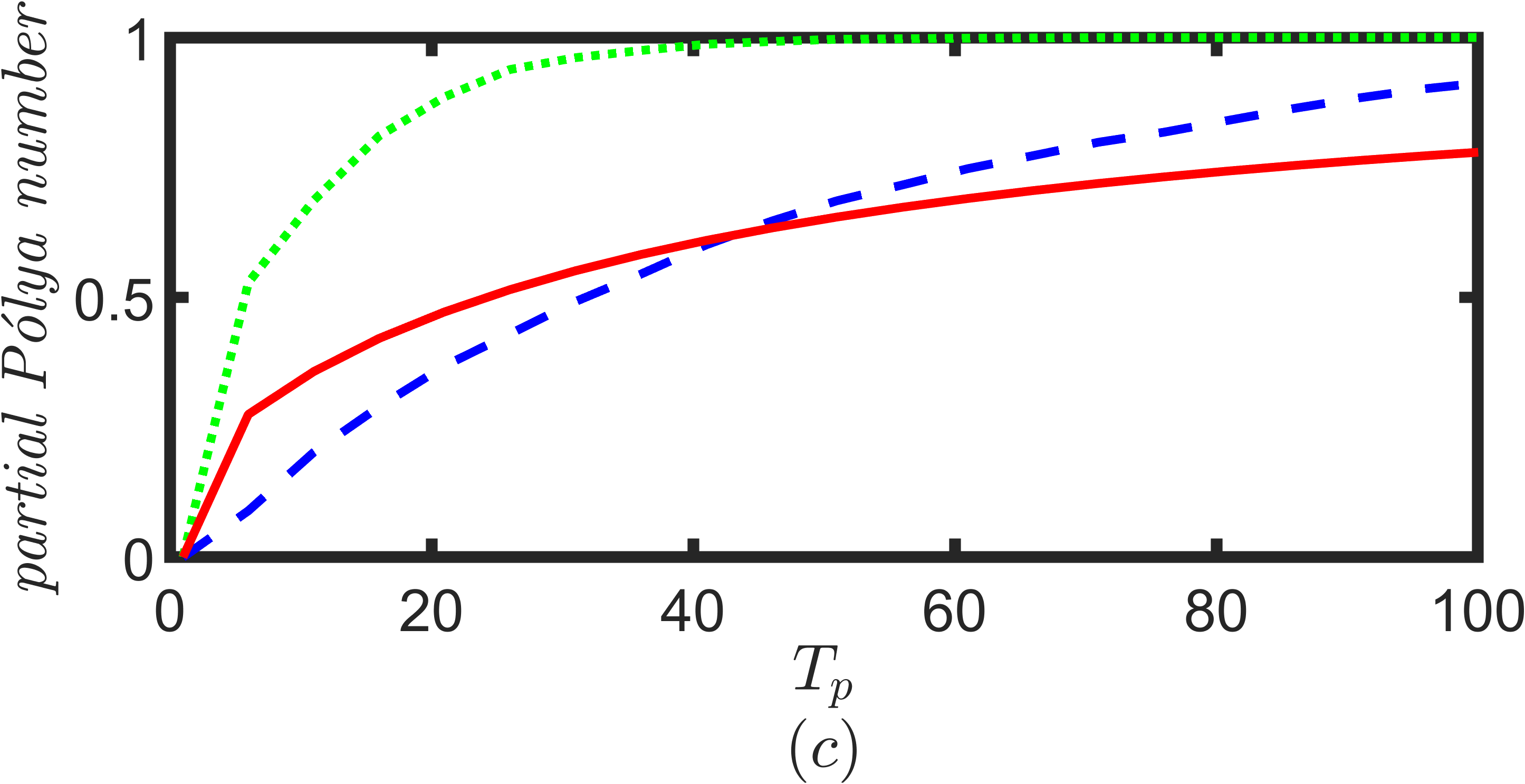} 
\includegraphics[width=2.4in,height=1.3in]{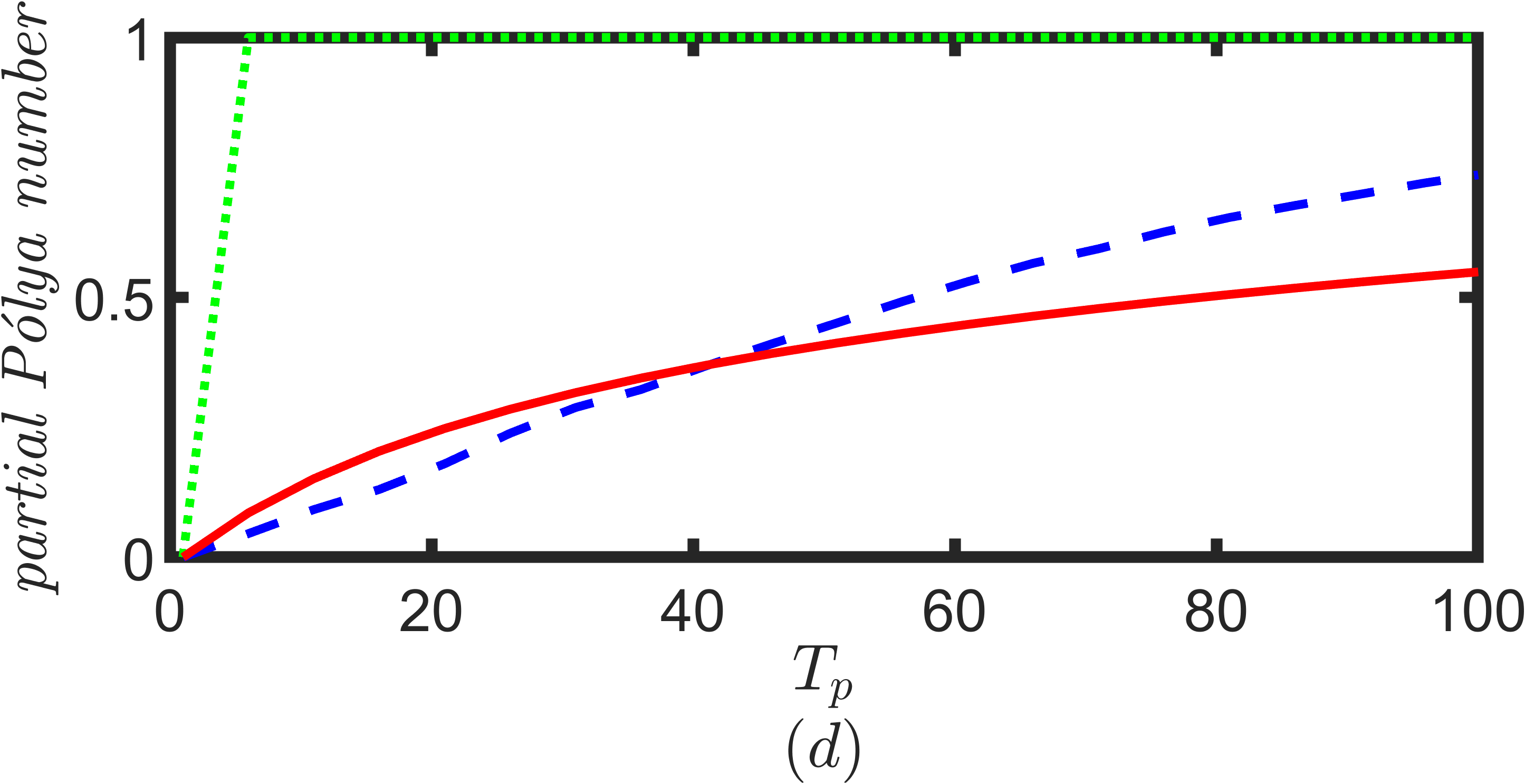} 
\includegraphics[width=2.4in,height=1.3in]{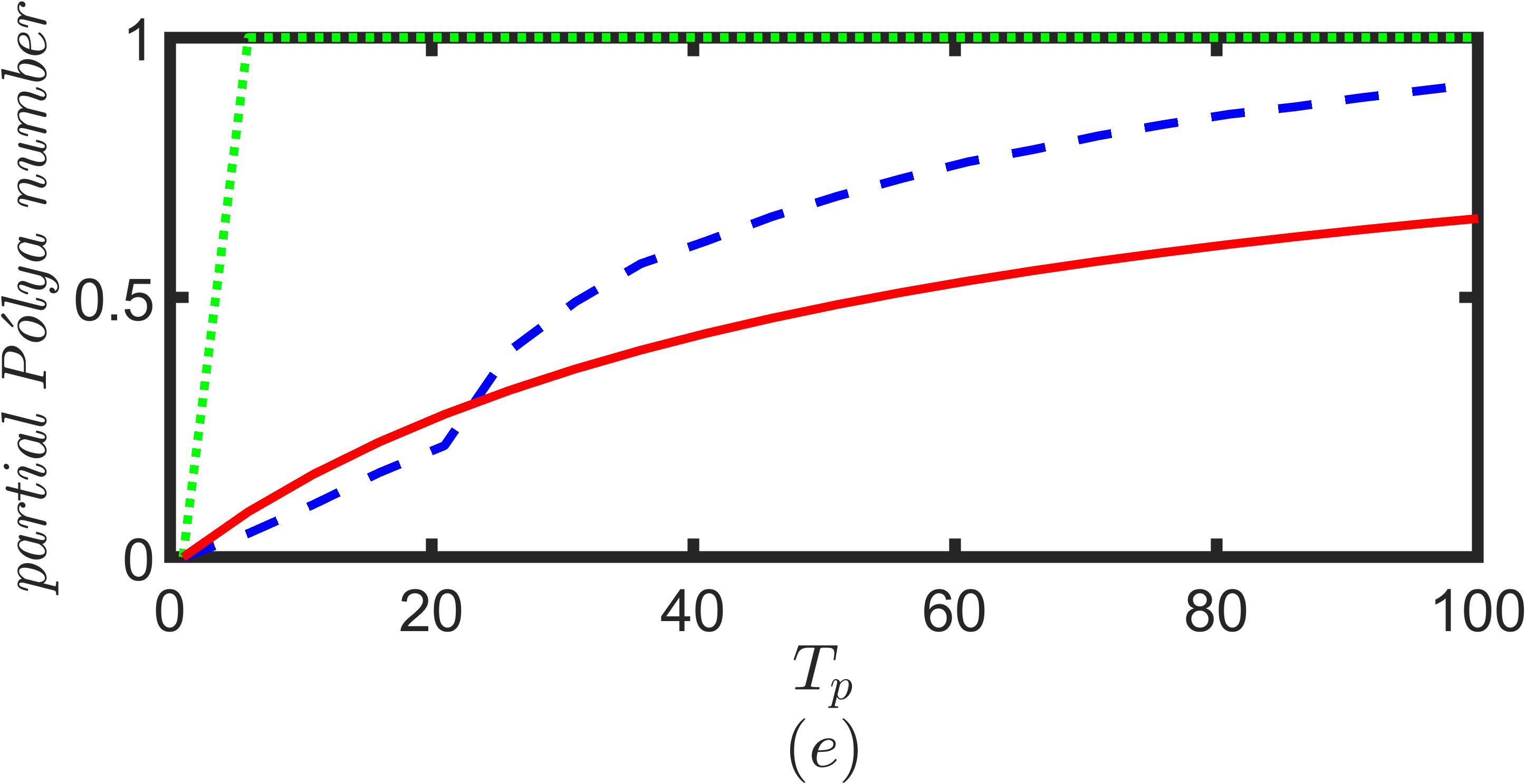} 
\includegraphics[width=2.4in,height=1.3in]{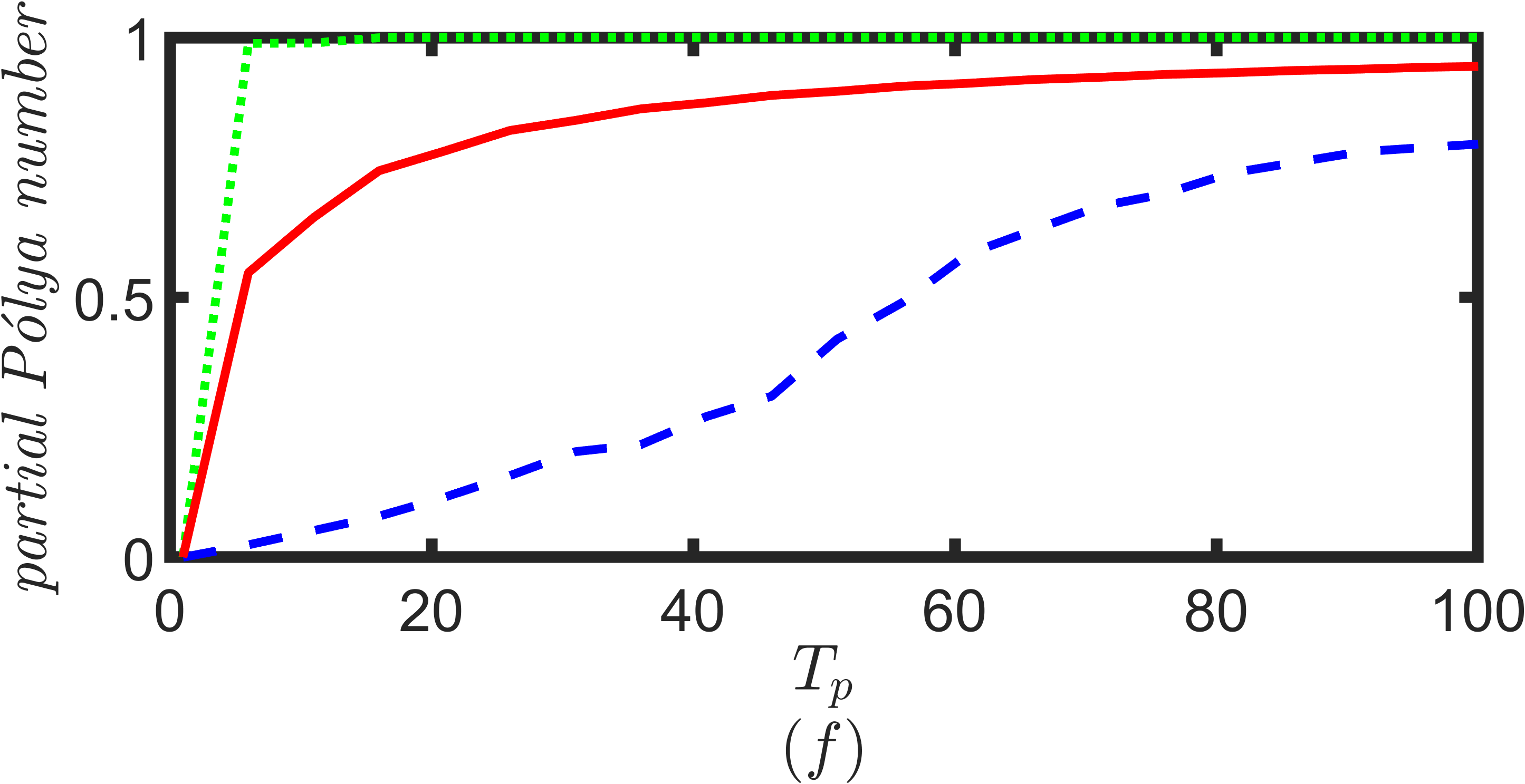} 
     \caption{Convergence of the partial P\'olya number for Grover (green dotted line), Fourier (blue dash line) and Classical (red solid line) walkers on six different two-layered multiplex networks, each consists of 100 nodes. The top and bottom layers of each multiplex network are constructed from combinations of scale-free (SF), complete (CP) and star networks with 50 nodes. (a) SF-SF (b) SF-CP (c) SF-STAR (d) CP-CP (e) CP-STAR and (f) STAR-STAR. For the case of SF-SF, two different scale-free networks are chosen. Moreover, the hub node of the star network is taken as the first node. Partial P\'olya number is calculated by choosing a set of finite time steps $T_p \in \{1,5,10,...,100\}$ with a gap of $5$ units. Each walk is initiated from the node 1 and for both Grover and Fourier walks, initial coin state is chosen as the uniform superposition of coin states.}
     \label{recurrence_big_network}
\end{center}
\end{figure}

In addition to the exploration on probability profiles, we have studied the recurrence probability of the Fourier, Grover and classical walkers on the six different multiplex networks. We calculate the partial P\'olya number for CRW, Grover and Fourier walks by choosing a set of finite time steps $T_p \in \{1,5,10,...,100\}$ with a gap of $5$ units. Our purpose is to make an estimation of the convergence of the P\'olya number for each walk. We initialize each walker from node 1 and for both Grover and Fourier walks, initial coin state is chosen as the uniform superposition of coin states. (That is, $|1\rangle_p \otimes \frac{1}{\sqrt{d_1}}\sum_{r=1}^{d_1}|f_1(r)\rangle_c$). Figure \ref{recurrence_big_network} shows the convergence of the partial P\'olya number for CRW, Grover and Fourier walks on the six different multiplex networks. From Figure \ref{recurrence_big_network}, one can identify that the Grover walk exhibits recurrence on most of the network structures studied here. On the other hand, Fourier and classical walkers show no recurrence within 100 time steps. For all the plots in Figure \ref{recurrence_big_network}, initially, the convergence of the partial P\'olya number of the Fourier walk is low compared to that of a classical walker. However, as time elapses, the convergence of the Fourier walk surpasses that of CRW except for the case of STAR-STAR multiplex network. 

 We apply the broken link decoherence model for the six different multiplex networks as well by following the same procedure described in section \ref{decoherence}. To realize the broken link decoherence model, we have removed some edges of the multiplex networks randomly and have calculated the average probability distribution of the QW after 100 time steps by averaging over 1000 trials. In section \ref{decoherence}, we observed the emergence of classical signature in the probability distribution even for a single broken link on the toy multilayer network. However, since the number of nodes in the six different multiplex networks are relatively large, we couldn't observe a fast convergence to the classical behaviour when a single edge is broken. However, when the number of broken links increases, the convergence to the classical distribution becomes faster. Hence, the impact of decoherence depends upon the number of broken edges. While our investigation has shed some light on decoherence on multilayer network, it is important to acknowledge that the scope of this research is not fully comprehensive. There remains significant potential for future studies to expand upon these findings to gain a deeper understanding of how decoherence impact on the propagation of the quantum walker on multilayer networks.
 
\section{Discussion}
In this paper we studied the dynamics of discrete-time quantum walks on multilayer networks. We derived recurrence formulae for the coefficients of the wave function of a quantum walker on an undirected graph with finite nodes. Then, by extending these formulae to include extra layers, we developed a simulation to mimic the evolution of the quantum walker on a multilayer network. While multilayer networks have been studied in the context of CTQWs, to the best of our knowledge, there is a lack of literature related to DTQWs on multilayer networks. Hence, the prime objective of this study was to present a comprehensive mathematical framework to model DTQWs on multilayer networks with the aim of bridging this gap. In this regard, we employed our mathematical model to analyze the time-averaged probability and the return probability of the quantum walker on multilayer networks in relation to Fourier and Grover walks. Moreover, we studied the impact of decoherence on the progression of Fourier walk on mulilayer networks. For sake of clarity and readability, first we used a toy muliayer network to conduct our analysis. Later we extended our analysis to much larger synthetic multilayer networks. Our study reveled that the Grover walk on a multilayer network exhibits rich dynamics. For instance, the Grover walker displays a periodic behaviour of occupying the top and bottom layers of a two-layered multiplex network constructed from a complete graph or a star graph. Further research and analysis are essential to unlock the full potential of these insights for practical quantum applications, e.g. for quantum computation and quantum communication. Moreover, in relation to the recurrence probability, the Grover walker returns to the initial position faster than both Fourier and Classical walkers. In the context of QWs, the recurrence probability has a deep link to the localization property, playing a pivotal role in diverse applications, including quantum search algorithms and topological insulators \cite{kiumi2022return}. Hence, there seems to be significant potential for future studies related to the return probability on multilayer networks. Another finding of this study is that the QWs on multilayer networks are vulnerable to decoherence arsing from randomly broken links. Multilayer networks with a smaller number of nodes are sensitive to defects even in a single edge. However, tolerance of the QWs for the decoherence arising from a defect in single edge increases as the number of nodes increases. Nonetheless, when there exists more defects in the edges, the probability distribution of the QWs converges to the classical distribution very quickly. As a future extension of this study, one could explore how other forms of decoherence models, like Pauli channels as well as amplitude and phase damping in the coin degree of freedom, make an impact on the QWs on multilayer networks. In summary, we anticipate that the mathematical analysis we have performed here may have a profound influence on a broad spectrum of problems which can be modelled or assisted by DTQWs on multilayer networks.

\section{Acknowledgement}
M. N. Jayakody acknowledges the President Scholarship Program at Bar-Ilan University. P. Pradhan acknowledges Science and Engineering Research Board (SERB) grant TAR/2022/000657, Govt. of India. This research was funded in part by the Israeli Innovation Authority under Project No. 73795, by the Pazy Foundation, by the Israeli Ministry of Science and Technology, and by the Quantum Science and Technology Program of the Israeli Council of Higher Education.

\bibliography{sample-base}

\appendix

\onecolumngrid

\section[\appendixname~\thesection]{Coin flip and Shift operation}\label{App2}
Let us represent the state vector $|\psi_t\rangle$ given by \eqref{State_vector_at_t} in the block matrix form. That is $|\psi_t\rangle \equiv {\bf N}_t$. Now let us apply the global coin operator $C$ given in \eqref{coin_opertor} on the total wave function $|\psi_t\rangle$ in \eqref{State_vector_at_t}. Then we get the following expression
\begin{equation}\label{C_on_state_vectorEq1}
    C|\psi_t\rangle=\sum_{x=1}^{n}\sum_{i=1}^{d_x} \biggl(\sum_{j=1}^{d_x} \alpha_{x,f_x(j)}(t)C^{(x)}_{ij}\biggr)|x\rangle_p| f_x(i) \rangle_c
\end{equation} 
Let us write $\tilde{\alpha}_{x,f_x(i)}(t)=\biggl(\sum_{j=1}^{d_x} \alpha_{x,f_x(j)}(t)C^{(x)}_{ij}\biggr)$. Then we can rewrite \eqref{C_on_state_vectorEq1} as follows
\begin{equation}\label{C_on_state_vectorEq2}
    C|\psi_t\rangle=\sum_{x=1}^{n}\sum_{i=1}^{d_x} \tilde{\alpha}_{x,f_x(i)}(t)|x\rangle_p| f_x(i) \rangle_c
\end{equation} 
Note that, $C|\psi_t\rangle$ can also be represented in the block matrix form. Let us denote $ C|\psi_t\rangle \equiv {\bf \widetilde{N}}_t $. Now let us apply the shift operator $S$ given in \eqref{shift_opertor} on $ C|\psi_t\rangle$. The result can be written as
\begin{equation}\label{SC_on_state_vectorEq3}
    SC|\psi_t\rangle=\sum_{x=1}^{n}\sum_{i=1}^{d_x} \tilde{\alpha}_{x,f_x(i)}(t)|f_x(i)\rangle_p|x \rangle_c
\end{equation} 
Let the block matrix form of $SC|\psi_t\rangle$ be  ${\bf {M}}_t$. By observing \eqref{C_on_state_vectorEq2} and \eqref{SC_on_state_vectorEq3} one can argue that the elements in the $x^{th}$ row of ${\bf \widetilde{N}}_t$ is equal to the elements in the $x^{th}$ column of ${\bf M}_t$. That is, ${\bf {M}}_t$ is equal to the  transpose of ${\bf \widetilde{N}}_t$. In addition, we have the following expression  
\begin{equation}\label{state_vectorEq4}
SC|\psi_t\rangle=|\psi_{t+1}\rangle=\sum_{x=1}^{n}\sum_{i=1}^{d_x} \alpha_{x,f_x(i)}(t+1)|x \rangle_p|f_x(i) \rangle_c
\end{equation} 
Thus, for each $x \in \{1, \ \hdots \ n \}$ and $r \in \{1, \ \hdots \ d_x \}$ we can derive a recurrence relationship between the elements of ${\bf N}_t$ and ${\bf N}_{t+1}$ as follows
\begin{equation}\label{state_vectorEq5}
    \alpha_{x,f_x(r)}(t+1)=\tilde{\alpha}_{f_x(r),x}(t)
\end{equation} 
This completes the proof. 

\section[\appendixname~\thesection]{Example of graph with four nodes}\label{App1}
\begin{figure}[h]
\begin{center}
\includegraphics[width=3.8in, height=2.5in]{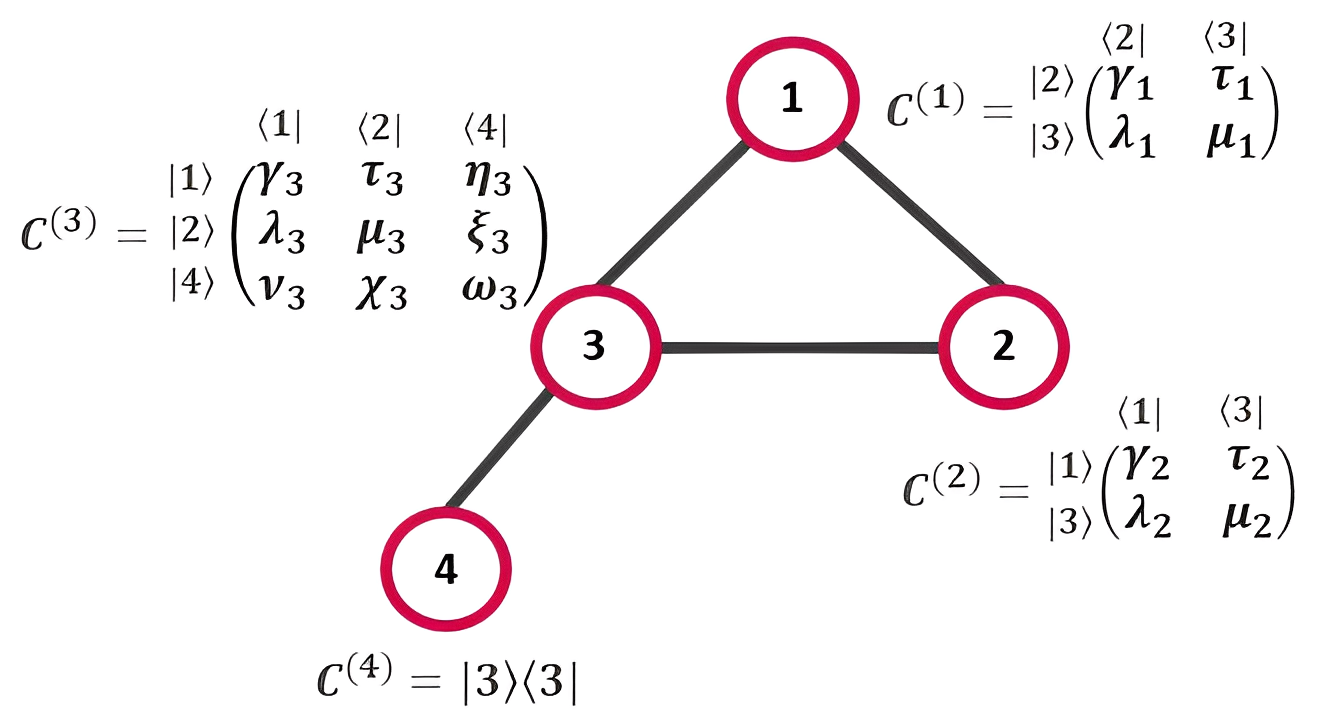}
\caption{Schematic diagram of a graph of four vertexes and the coin operators associated with each vertex. For each vertex $x$, the elements of the coin operator associated to $x$ are chosen in such a way that $C^{(x)}(C^{(x)})^{\dagger}=(C^{(x)})^{\dagger}C^{(x)}=\mathbb{I}$ where $x \in \{1 ,2,3,4\}$}
\label{regular_graph_of_four_vertexes}
\end{center}
\end{figure}

Let us consider the graph given in Figure \ref{regular_graph_of_four_vertexes}. The corresponding set of vertices and edges can be written as $V=\{|1\rangle_p,|2\rangle_p,|3\rangle_p,|4\rangle_p\}$ and $E=\{\{1,2\},\{1,3\},\{2,1\},\{2,3\}, \{3,1\},\{3,2\},\{3,4\},\{4,3\} \}$ respectively. 

\noindent Now let us define the sets of $\mathcal{B}_x$ as follows; $\mathcal{B}_1=\{2, \ 3\}$, $\mathcal{B}_2=\{1, \ 3\}$, $\mathcal{B}_3=\{1, \ 2, \ 4\}$ and $\mathcal{B}_4=\{3 \}$. The function $f_x(r)$ ($r^{th}$ element of the $\mathcal{B}_x$ set) can be defined as follows; $f_1(1)=2, \ f_1(2)=3$; $f_2(1)=1, \ f_2(2)=3$; $f_3(1)=1, \ f_3(2)=2, \ f_3(3)=4$  and $f_4(1)=3$. Then, the state vector of the quantum walker at time $t$ can be written as
\begin{equation}\label{State_vector_regular_graph_of_four_vertexes}
|\psi_t\rangle=\sum_{x=1}^{4}\sum_{r=1}^{d_x}\alpha_{x,f_x(r)}(t)|x\rangle_p |f_x(r) \rangle_c
\end{equation} 
The state vector given in \eqref{State_vector_regular_graph_of_four_vertexes} can be represented in terms of the block matrix ${\bf N}_t$ as
\begin{equation}\label{block_matrix_A_graph_of_four_vertexes}
 {\bf N}_t= \bordermatrix{ & |1 \rangle_c & |2 \rangle_c & |3 \rangle_c & |4 \rangle_c \cr 
  |1\rangle_p  & 0 & \alpha_{1,2}(t) & \alpha_{1,3}(t) & 0 \cr 
  |2\rangle_p  & \alpha_{2,1}(t) & 0 & \alpha_{2,3}(t) & 0 \cr  
  |3\rangle_p  & \alpha_{3,1}(t) & \alpha_{3,2}(t) & 0 & \alpha_{3,4}(t) \cr 
  |4\rangle_p  & 0 & 0 & \alpha_{4,3}(t) & 0 \cr} 
\end{equation}
Note that, by updating the elements of ${\bf N}_t$ using the relationships given in \eqref{update_rule1} and \eqref{update_rule2} one can mimic the evolution of a QW on the graph in Figure \ref{regular_graph_of_four_vertexes}.

\section[\appendixname~\thesection]{Probability of finding the classical walker on top and bottom layers of six different multiplex networks }\label{App3}
The following set of graphs shows the probability of finding the classical walker on six different synthetic two-layered multiplex networks with 100 nodes.
\begin{figure}[h] 
\begin{center}
    \includegraphics[width=2.3in,height=1.3in]{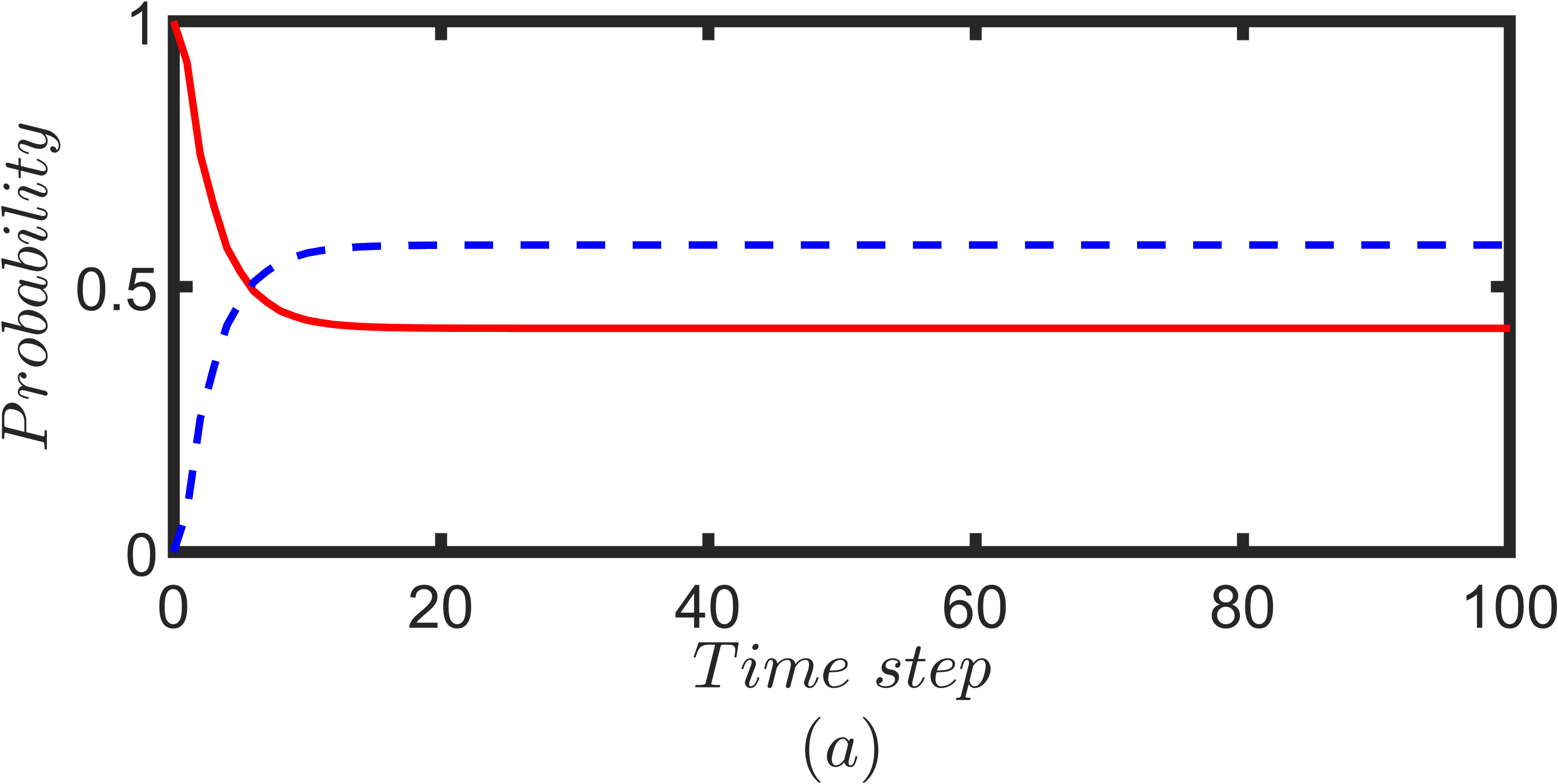} 
        \includegraphics[width=2.3in, height=1.3in]{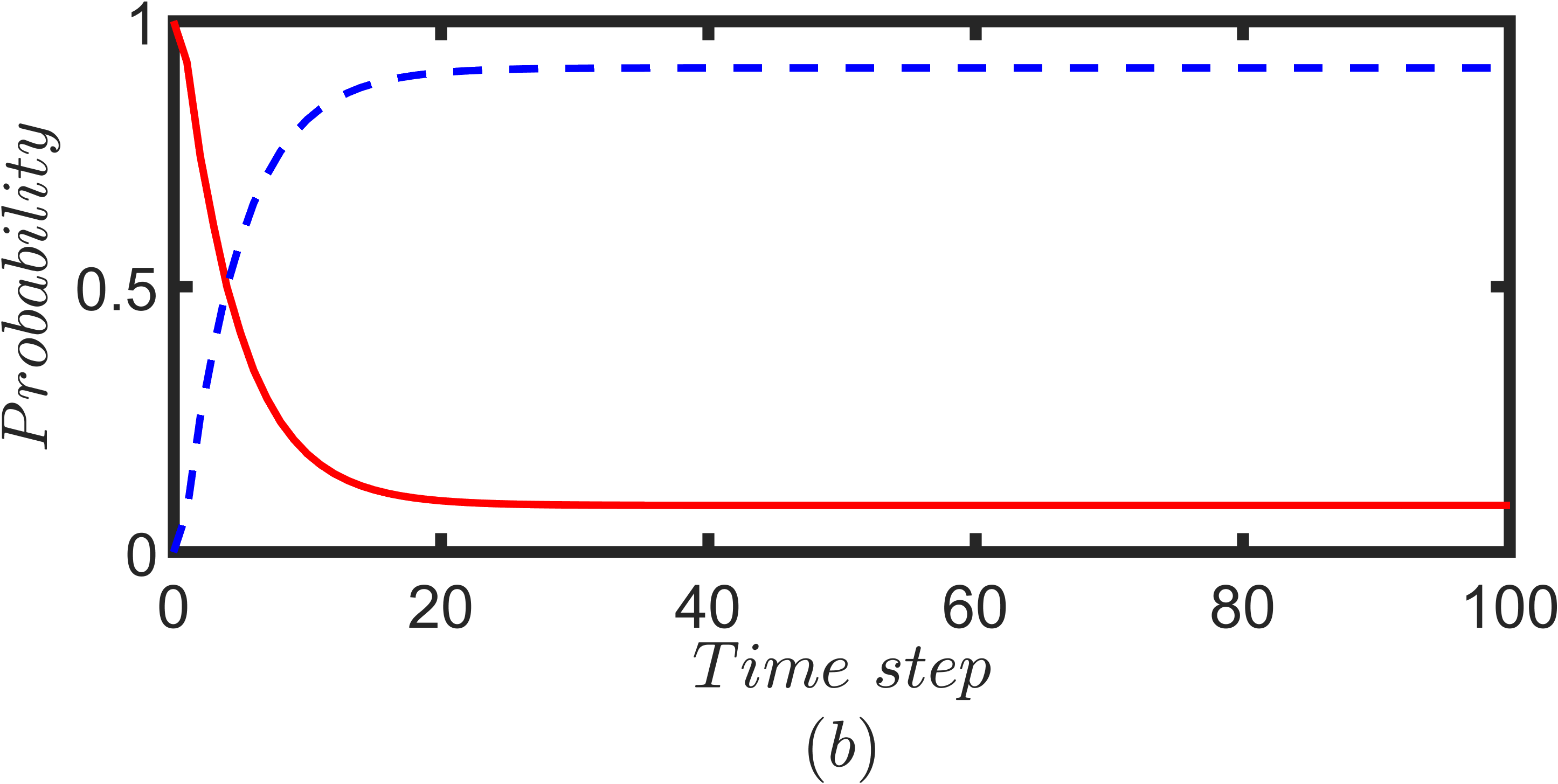} 
        \includegraphics[width=2.3in, height=1.3in]{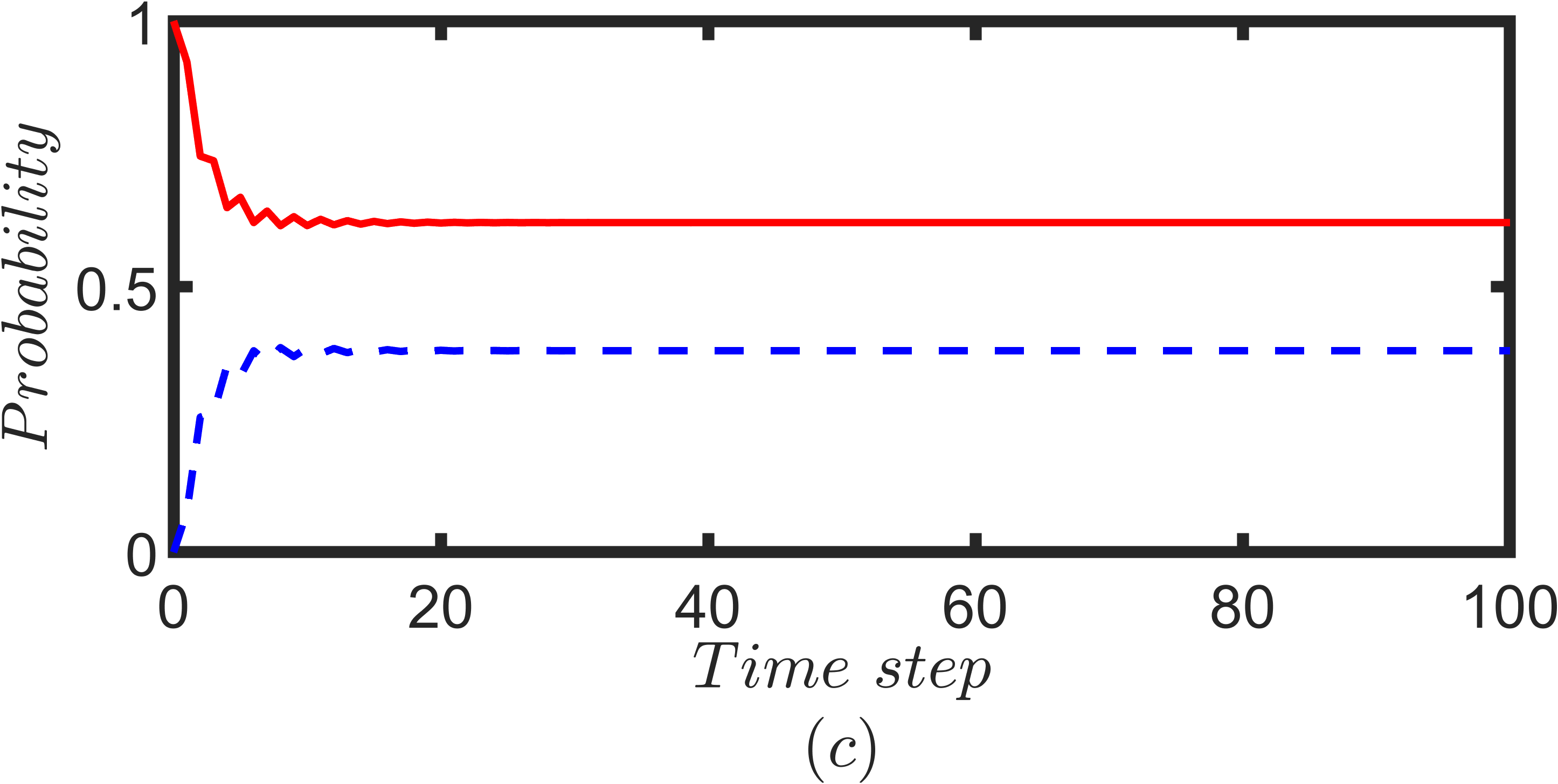} 
        \includegraphics[width=2.3in, height=1.3in]{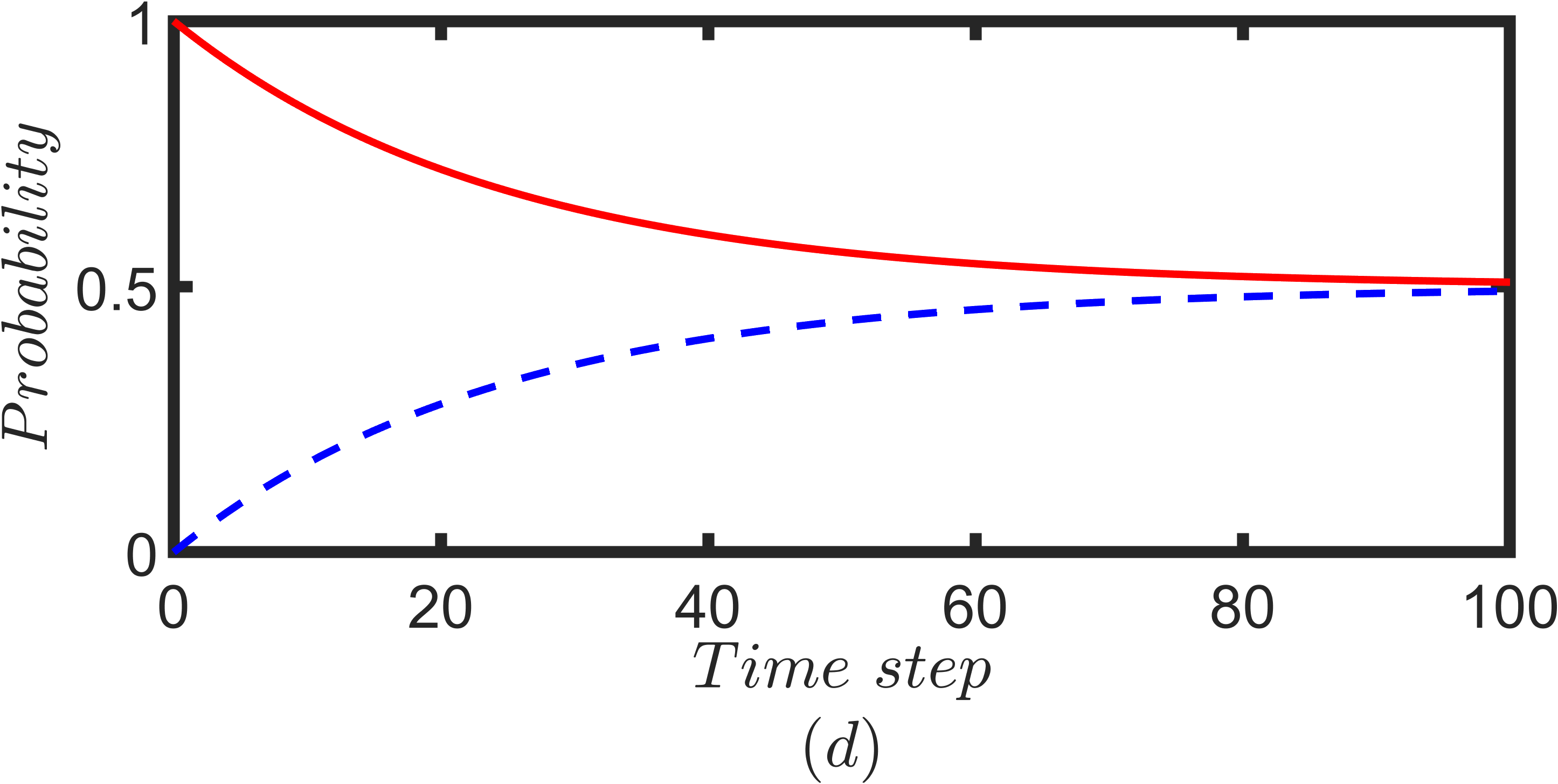}
        \includegraphics[width=2.3in, height=1.3in]{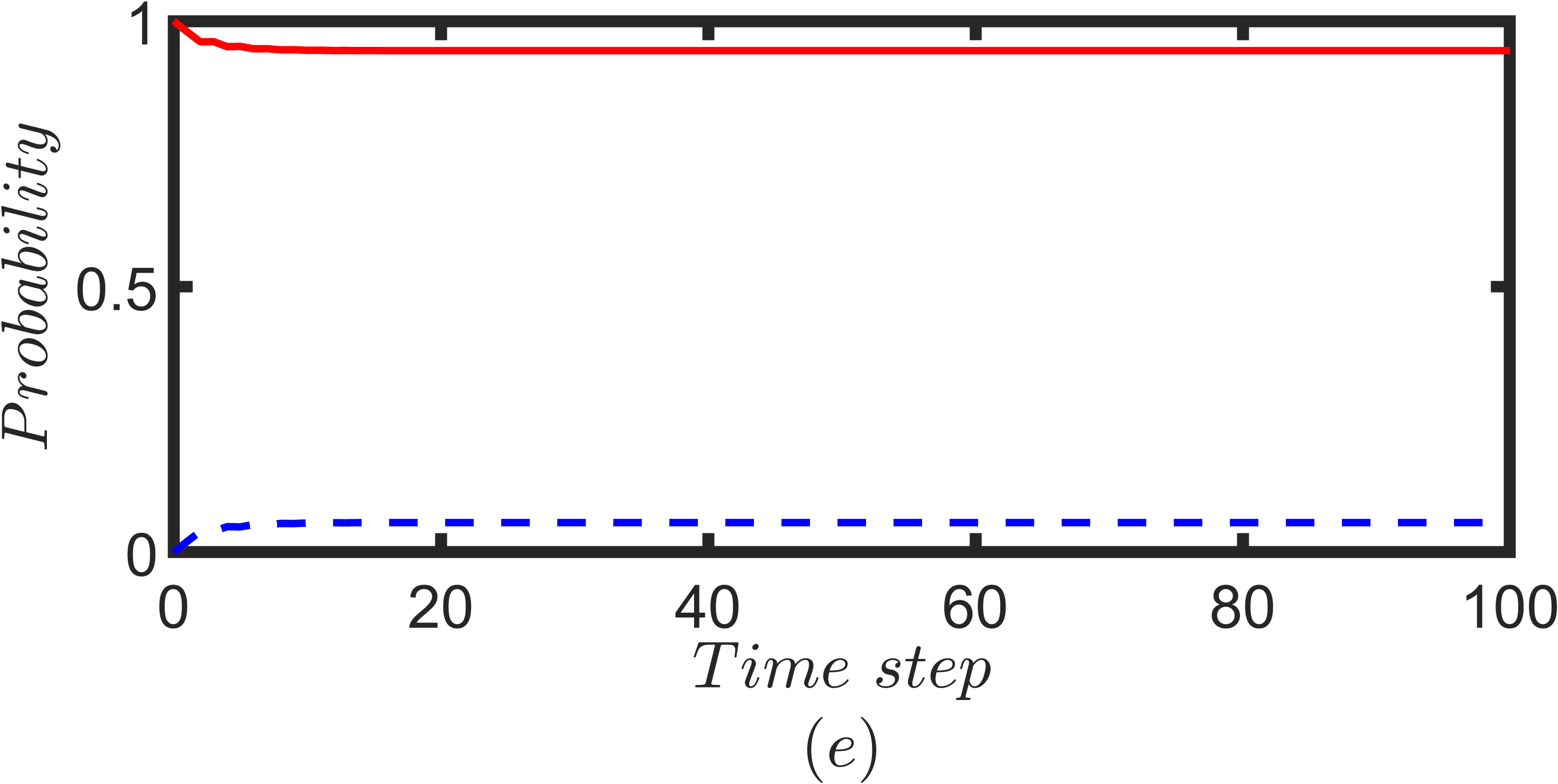}
        \includegraphics[width=2.3in, height=1.3in]{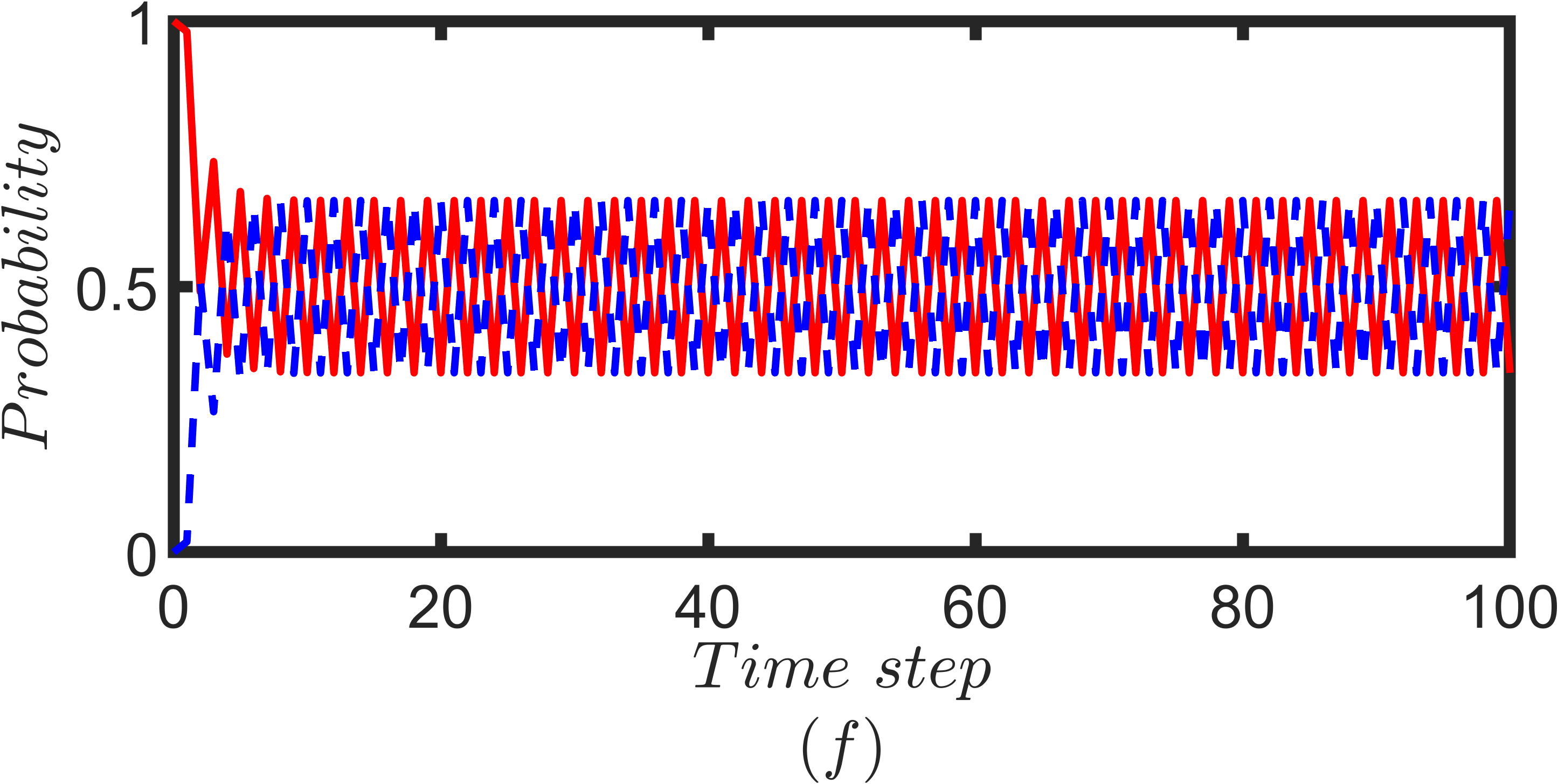}
     \caption{This figure illustrates the probability of finding the classical walker on the top layer (red solid line) and the bottom layer (blue dotted line) of six different two-layered multiplex networks, each consists of 100 nodes. The top and bottom layers of each multiplex network are constructed from combinations of scale-free (SF), complete (CP) and star networks with 50 nodes. (a) SF-SF (b) SF-CP (c) SF-STAR (d) CP-CP (e) CP-STAR and (f) STAR-STAR. For the case of SF-SF, two different scale-free networks are chosen. Moreover, the hub node of the star network is taken as the first node. CRW is initiated from vertex 1 and for each time step up to 100 steps, the probability of finding the walker on a given layer is calculated by summing the probabilities of finding the walker at each node corresponding to that layer.}
\label{Prob_on_layers_CRW_big_network}
\end{center}
\end{figure}

\end{document}